\documentclass[fleqn,authoryear,12pt]{elsarticle}
\usepackage[authoryear]{natbib}
\usepackage{setspace}
\usepackage{soul,xcolor}
\usepackage{graphicx,epstopdf}
\usepackage{enumitem}
\usepackage{color}
\usepackage{amsfonts}
\usepackage{amssymb}
\usepackage{booktabs} % for better table lines
\usepackage{arydshln} % for dashed lines in tables
\usepackage{array} % for table styling
\usepackage{caption} % for table caption
\usepackage{mathrsfs}
\usepackage{empheq}

\DeclareGraphicsExtensions{.eps}
\usepackage{subcaption}
\usepackage{color}
\usepackage[tmargin = 0.8 in, bmargin = 0.8 in, lmargin = 0.6 in, rmargin =0.6 in ]{geometry}
\usepackage{amsmath}
\usepackage{bm}
\usepackage{relsize}
\usepackage{amsthm}
\usepackage{stmaryrd}
\usepackage{titlesec}
\usepackage{array}
\usepackage[most]{tcolorbox}
\usepackage[colorlinks,
linkcolor=blue,
anchorcolor=blue,
citecolor=blue]{hyperref}
\usepackage[nameinlink]{cleveref}
\crefname{ineq}{Inequality}{Inequalities}
\creflabelformat{ineq}{#2{\upshape(#1)}#3}
\usepackage[nameinlink]{cleveref}
\crefname{ineq}{Inequality}{Inequalities}
\creflabelformat{ineq}{#2{\upshape(#1)}#3} 
\usepackage{placeins} % do not use [section]: fewer forced float barriers $\rightarrow$ less empty vertical space
\usepackage{stackengine}
\usepackage[normalem]{ulem}
\usepackage{amsmath}
\usepackage{amsthm}

\newcommand{\eqn}[1]{{Eq.\,(\ref{#1})}}

\newcommand{\pd}[2]{\frac{\partial{#1}}{\partial{#2}}}

\newcommand{\md}{\mathrm{d}}

\newcommand{\E}{\psi_\mathrm{elastic}}
\newcommand{\bme}{\boldsymbol{\varepsilon}}

\newcommand{\bmsi}{\boldsymbol{\sigma}}

\newcommand{\bmn}{\boldsymbol{n}}

\newcommand{\bms}{\boldsymbol{s}}
\newcommand{\Gc}{G_{\rm c}}

\newcommand{\me}{\mathrm{meso}}

\usepackage{accents}
\newcommand*{\dt}[1]{%
	\dot{#1}}
\newcommand{\xbar}[1]{%
	\hbox{%
		\vbox{%
			\hrule height 0.5pt % The actual bar
			\kern 0.5ex%         % Distance between bar and symbol
			\hbox{%
				\kern-0.1em%      % Shortening on the left side
				\ensuremath{#1}%
				\kern-0.1em%      % Shortening on the right side
			}%
		}%
	}%
} 
\theoremstyle{remark}
\theoremstyle{break}
\newtheorem{remark}{Remark}
\newtheorem{remark*}{Remark}
\theoremstyle{definition}

\begin{document}
\raggedbottom % avoid \flushbottom stretching huge gaps when floats land on otherwise short pages
\setlength{\textfloatsep}{10pt plus 4pt minus 4pt}
\setlength{\floatsep}{10pt plus 2pt minus 2pt}
\setlength{\intextsep}{10pt plus 2pt minus 2pt}
    \setstcolor{red}
    \begin{frontmatter}
\title{\large A unified sharp-diffusive phase-field model for bulk and interfacial cohesive fracture}

\author[NWPU]{Ye-Hang~Qin}

% 将两位通讯作者的 \corref 都设置为 cor1
\author[NWPU,Lab]{Ye~Feng\corref{cor1}}
\ead{fengye@nwpu.edu.cn}

%\author[Tongji,Lab2]{Yan~Li\corref{cor1}}
%\ead{liyan@tongji.edu.cn}

\address[NWPU]{School of Aeronautics, Northwestern Polytechnical University, Xi'an 710072, China}
\address[Lab]{National Key Laboratory of Strength and Structural Integrity, Xi'an 710072, China}

% 翻译为英文并补全了上海和邮编
%\address[Tongji]{School of Aerospace Engineering and Applied Mechanics, Tongji University, Shanghai 200092, China}
%\address[Lab2]{Key Laboratory of Design and Manufacture of Composite Materials, Ministry of Education, Shanghai 200092, China}

% 对应的脚注文本也只需要写一个 cor1
\cortext[cor1]{Corresponding author.}
\begin{abstract}
In traditional phase-field modeling of multiphase materials, a significant challenge arises from the non-local nature of fracture energy regularization, where interfacial toughness is inherently coupled with the properties of the surrounding bulk phases. Achieving consistency with prescribed material properties typically necessitates complex corrections and exceptionally fine local mesh refinement near the interfaces. To address this fundamental issue, we leverage the capacity of the recently proposed $\Omega^2$-model  to manifest Dirac-like damage concentration and emergent displacement discontinuities, while introducing \emph{an analytical, strongly localized  interfacial source term} $q_{\phi}$ into the phase-field formulation. It should be emphasized that the ``sharp" nature of the proposed model manifests as a naturally emergent strong discontinuity within a continuum framework, fundamentally distinguishing it from inherently discrete approaches such as cohesive element method. This allows for the independent and precise control of interface toughness in a straightforward manner. Theoretical analysis further reveals that the proposed framework can describe the cohesive failure of both bulk and interfacial regions using a \emph{unified} set of parametric equations for the cohesive law, where the model parameters are directly determined by the local material properties without the need for additional corrections. The model's versatility is numerically validated through a series of benchmarks. The results confirm that the proposed model not only accurately reproduces diverse interfacial cohesive laws but also captures the intricate competition between interfacial debonding and matrix cracking. This sharp-diffusive phase-field model may provide a robust and computationally efficient tool for predicting complex fracture trajectories in sophisticated engineering materials.
\end{abstract}
    \end{frontmatter}
    \section{Introduction}
   Multiphase materials, such as fiber-reinforced composites, multi-phase alloys, and porous ceramics, are widely used in aerospace, automotive, and energy sectors due to their exceptional specific strength and tailorable mechanical properties. By carefully designing the constituent phases, their volume fractions, and spatial distributions, these materials can achieve superior performance unattainable by single-phase materials. However, their heterogeneous microstructures often lead to complex failure behaviors, which pose a significant challenge for high-fidelity numerical simulations.
   Among various failure modes, interfacial debonding is particularly critical, as the interfaces between different phases are typically the weakest links, governing the overall structural integrity. Moreover, the interaction between matrix cracking and interfacial fracture can induce complex failure mechanisms, such as crack deflection and penetration, phenomena that are also difficult to capture accurately with traditional fracture models.
   
   In recent decades, the phase-field method has emerged as a prominent tool for simulating fracture in solids. Its theoretical foundation can be traced back to the variational approach to fracture proposed by \cite{francfort1998revisiting}, which interprets crack evolution as a spontaneous process of system energy minimization. Subsequently, \cite{bourdin2000numerical} enabled its numerical solution via Ambrosio-Tortorelli regularization, \cite{karma2001phase} independently developed a phase-field description for dynamic fracture based on Ginzburg-Landau theory, and 
   \cite{miehe2010thermodynamically} established a thermodynamically consistent phase-field framework, laying a systematic foundation for the geometric interpretation of crack density and its finite element implementation. 
   
To capture the complex failure mechanisms in heterogeneous and composite materials, the phase-field fracture framework has been extensively developed. A pivotal advancement was made by \cite{nguyen2016phase}, who elegantly modeled the intricate interaction between bulk crack propagation and interfacial damage within realistic microstructures. To explicitly differentiate distinct failure modes, \cite{bleyer2018phase} proposed a multi-phase-field formulation that assigns separate damage variables to interacting mechanisms in anisotropic brittle materials. This anisotropic conceptualization was further advanced by \cite{zhang2019phase}, who formulated a phase-field model driven by anisotropic energy decomposition to characterize fracture in fiber-reinforced composite laminates.
 Pushing the boundaries of coupled failure, \cite{yue2025triple} introduced a sophisticated triple-damage model that integrates reduced-order homogenization to concurrently capture intra-laminar and inter-laminar degradation.

In the phase-field modeling of multiphase materials, a major challenge stems from the non-local nature of fracture energy regularization, which intrinsically couples the interfacial toughness with the properties of the surrounding bulk phases~\citep{hansen2019phase}. Consequently, achieving consistency with prescribed material parameters generally requires complex constitutive mapping or ad hoc corrections. This issue of non-locality has also been highlighted in the context of heterogeneous materials with continuously varying properties~\citep{vicentini2023phase}
  
  To mitigate this issue, \cite{hansen2019phase} introduced  a compensation method to address the length-scale competition between the crack phase-field and the interface region. Building upon this concept, a major contribution was made by \cite{yoshioka2021variational}, who derived an analytical expression for the effective fracture toughness of a diffuse interface. This analytical treatment enabled accurate simulations of interfacial crack propagation without requiring structural modifications to existing phase-field frameworks. In a parallel effort to decouple interface responses, \cite{bian2021novel,bian2024unified} proposed an interfacial element based phase-field-based cohesive zone model sharing phase-field degrees of freedom, achieving independent control over the normal and tangential properties of the interface. 
  Besides, \cite{bian2025adaptive} introduced a unified phase-field cohesive zone model capable of capturing both mixed-mode interfacial and matrix fracture. This model incorporates the directional energy decomposition proposed by \cite{feng2023unified}, employs a mixed-mode fracture criterion \citep{bian2025variationally}, and an adaptive mesh refinement strategy to significantly enhance computational efficiency.
  
 Despite the remarkable progress achieved by these aforementioned methods, a fundamental limitation persists: the precise prescription of interfacial fracture toughness in these approaches intrinsically depends on the assumption of a finite diffusive interface width. Consequently, to accurately resolve the interfacial degradation, these methods impose stringent mesh refinement requirements near the interface region, which substantially hinders the overall computational efficiency. 
 
 To overcome these limitations, we draw inspiration from the recently proposed $\Omega^2$-model \citep{feng2025JMPS}. This framework introduces an independent damage variable $\omega$, thereby relieving the phase-field variable $\phi$ from the dual burden of simultaneously describing crack geometry and stiffness degradation. A defining characteristic of the $\Omega^2$-model is that the damage field $\omega$ exhibits a robust Dirac-like concentration, which naturally manifests emergent strong displacement discontinuities. Meanwhile, the phase field retains its role in energetic regularization, effectively suppressing the spurious mesh sensitivity that typically plagues traditional local damage models. This strong localization property of $\omega$ provides the theoretical foundation for the present work, enabling the accurate resolution of interfacial failure using only a single layer of elements.
 
Building upon the $\Omega^2$-model, this paper proposes a unified sharp-diffusive phase-field framework for bulk and interfacial cohesive fracture. By incorporating a strongly localized analytical interfacial source term, $q_\phi$, into the formulation, the proposed framework reduces the effective interfacial fracture energy in a precisely controlled manner. Crucially, the ``sharp'' nature of this approach manifests as a naturally emergent strong discontinuity within a continuum framework, fundamentally distinguishing it from inherently discrete techniques such as cohesive zone elements. As will be demonstrated through theoretical and numerical analyses, the new model endows interfaces with independently specified fracture toughness and strength. Furthermore, it facilitates a unified representation of tensile and shear cohesive failures—across both bulk and interfacial regions—by directly integrating the parameterized cohesive laws of Feng and Li \citep{feng2021endowing,feng2022phase,feng2022phase_dissipation,feng2023unified}. Within this framework, model parameters are explicitly determined by local physical properties, eliminating the need for complex corrections. Finally, leveraging the intense localization of the damage field $\omega$, the proposed model is capable of capturing interfacial discontinuities using only a single layer of elements, thereby offering a significant enhancement in overall computational efficiency.
   
The remainder of this paper is organized as follows. Section~\ref{sec:1D} introduces the one-dimensional formulation of the proposed unified framework, detailing the energy functional, governing equations, and the derivation of interfacial cohesive laws both with and without the localized source term $q_{\phi}$. Section~\ref{sec:3D} generalizes this theoretical framework to multiple dimensions. Section~\ref{sec:fracture simulation} systematically validates the robustness and accuracy of the model through four representative numerical benchmarks. Finally, Section~\ref{sec:conclusion} summarizes the main contributions of this work.

    \section{The $ \Omega^2 $ phase-field model for fractures in multiphase materials: 1D case}\label{sec:1D} 
	\subsection{1D  energy functional}
\begin{figure}[!htbp] 
	\centering
	\includegraphics[width=0.5\linewidth]{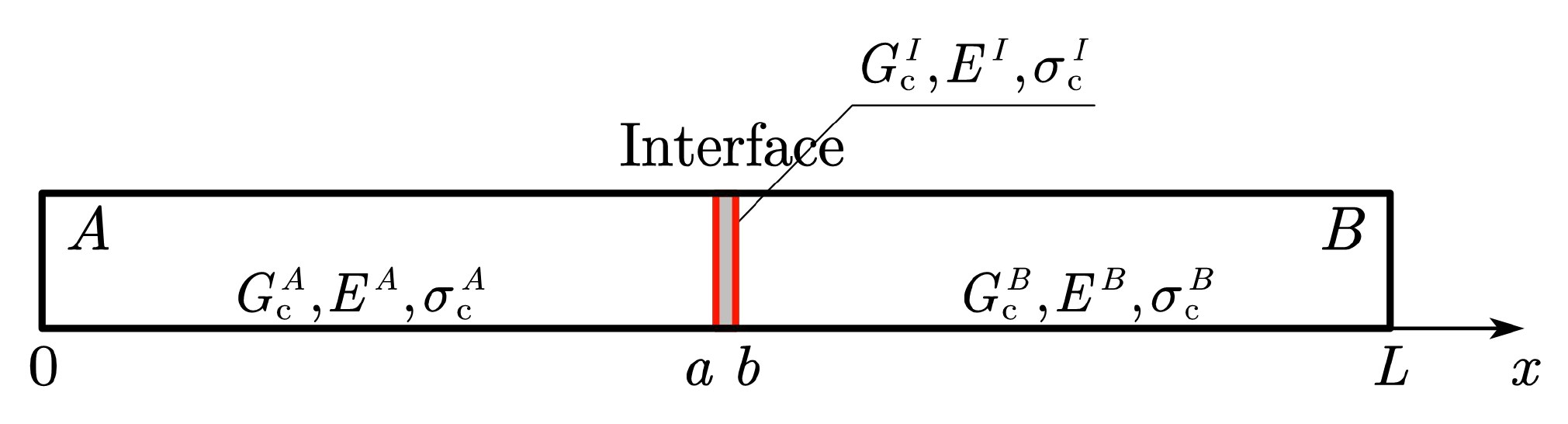}
	\caption{Schematic diagram of the 1D multiphase material}
	\label{fig:1D_bar}
\end{figure}

Consider the one-dimensional (1D) bar depicted in Fig.~\ref{fig:1D_bar}. The system comprises two constituent materials, $A$ and $B$, and an intermediate interface $I$, occupying the intervals $\mathcal{B}^{A}=(0,a)$, $\mathcal{B}^{B}=(b,L)$, and $\mathcal{B}^{I}=(a,b)$, respectively. The interface is initially modeled as a thin layer of width $h=b-a$; however, the sharp interface limit ($h \to 0^+$) will be theoretically examined in subsequent sections.

As illustrated in Fig.~\ref{fig:1D_bar}, the superscript $j \in \{A, B, I\}$ is used to designate the parameters associated with material $A$, material $B$, and the interface $ I $, respectively. For instance, the fracture energy, Young's modulus, and tensile strength of the interface $ I $ are represented by $G_\mathrm{c}^{I}$, $E^{I}$, and $\sigma_\mathrm{c}^{I}$, respectively.

In the $ \Omega^2 $-model, the total potential energy of the solid, $\mathcal{E}_\ell  $, is defined as a functional of the displacement field $ u $, the phase-field $ \phi\in[0,1] $, and a new damage field $ \omega\in[0,+\infty) $:
\begin{equation}\label{eq:energy functional}
	\mathcal{E}_\ell[u,\phi,\omega]= \int_{\mathcal{B}} \Big[\psi_{\mathrm{crack}}\left( \phi ,\partial_x \phi ,\omega \right) +\E\left( {\varepsilon },\omega \right)\Big]\md V,
\end{equation}
where the external force potential is omitted for brevity, $ \psi_{\mathrm{crack}} $ is the crack energy density, defined as
\begin{equation}\label{eq:phi_crack}
	\psi_{\mathrm{crack}}=\frac{G_{\mathrm{c}}}{2\ell}\Big[\Omega(\phi)^2 +\ell ^2\left(\partial_x\phi\right)^2 \Big] +\frac{G_{\mathrm{c}}}{\ell}(1-\phi)^2\omega,
\end{equation}
 and $ \psi_\mathrm{elastic} $ is the elastic strain energy:
\begin{equation}\label{eq:phi_elastic}
	\psi_\mathrm{elastic}=\frac{1}{2}E\varepsilon^2-\left( 1-g_1\left( \omega \right) \right) \frac{\left< \bar{\sigma} \right>^2 }{2E}.
\end{equation}
 Above, $ \Omega(\phi) $ is a characteristic function determined by the material's cohesive law, as demonstrated in \eqn{eq:Omega_phi}, $ \varepsilon=\partial_x u $ denotes the infinitesimal strain, $ \bar{\sigma}=E\varepsilon $ is the nominal (undamaged) stress, and $ \langle\cdot\rangle:=\max\{\cdot,0\} $ represents the Macaulay brackets. Note that the material parameters $ \Gc(x) $, $ E(x) $, and $ \sigma_{\mathrm{c}}(x) $ in our problem depend on the spatial position $ x $, as illustrated in Fig.~\ref{fig:1D_bar}, whereas the phase-field length $ \ell $ is set as a constant within $ \mathcal{B}=\cup_{j\in\{A,B,I\}} \,\mathcal{B}^j $. 
 
Furthermore, the mode-I energy degradation function, $g_1$, is defined as:
 \begin{equation}\label{eq:g_1}
 	g_1(\omega) = \frac{1}{1+c_1\omega}
 \end{equation}
where $c_1$ represents the ratio between the characteristic length of the mode-I cohesive zone ($\ell^\mathrm{ch}_1$) and the phase-field length parameter $\ell$:
 \begin{equation}\label{eq:c_1}
 	c_1 =\frac{\ell^\mathrm{ch}_1}{\ell} \quad \text{with}\quad \ell^\mathrm{ch}_1 =\frac{2G_\mathrm{c}E}{\sigma_\mathrm{c}^2} .
 \end{equation}
As evident from \eqn{eq:phi_elastic}, the physical meaning of $\omega$ becomes clear: $\omega=0$ corresponds to an intact state without any degradation, whereas $\omega=+\infty$ implies that the material is completely fractured.

For a homogeneous material, as proved by \cite{feng2025JMPS} in 1D and \cite{feng_Li2026} in multiple dimensions, the $ \Omega^2 $-model's cohesive law (traction--separation law) is parameterized by
\begin{equation}\label{eq:Omega_phi}
	\hat{\sigma} = 1 - \phi^*\quad\text{and}\quad \hat{\varDelta} = \Omega(\phi^*),
\end{equation}
where  $\hat{\sigma}=\sigma/\sigma_\mathrm{c}$ and $\hat{\varDelta}=\varDelta/\varDelta_\mathrm{ch}$ represent the cohesive stress and crack opening displacement nondimensionalized by the tensile strength $\sigma_\mathrm{c}$ and the characteristic crack opening displacement $\varDelta_\mathrm{ch}=2G_\mathrm{c}/\sigma_\mathrm{c}$, respectively. 
\eqn{eq:Omega_phi} implies that the function $ \Omega(\cdot) $ in \eqn{eq:phi_crack} determines the shape of  traction-separation relation.

We list some widely used cohesive laws below. For other forms of cohesive laws and their parametric representations, we refer to \citep{feng2021endowing}.
\begin{itemize}
\item Linear softening law
\begin{equation}\label{eq:linear}
	\hat{\sigma} = 1 - \phi^*\quad\text{and}\quad \hat{\varDelta} = \Omega_\mathrm{linear}(\phi^*) = \phi^*
\end{equation}
\item Exponential softening law
\begin{equation}\label{eq:lne}
	\hat{\sigma} = 1 - \phi^*\quad\text{and}\quad \hat{\varDelta} = \Omega_\mathrm{exp}(\phi^*) = -\frac{1}{2}\ln (1-\phi^*) 
\end{equation}
\end{itemize}

Moreover, for convexity of energy and physical constraints of cohesive law, the $\Omega^2$-model imposes the following criteria on the function $\Omega(\phi)$ \citep{feng2025JMPS}:
\begin{equation}\label{eq:Omega_phi_dao}
\Omega(0)=\Omega^\prime(0)=0, \quad  \Omega^\prime(\phi)\geq 0,\quad \text{and}\quad  [\Omega(\phi)^2]^{\prime\prime}\geq0.
\end{equation}
\begin{remark}
	It should be noted that these parametric equations \eqref{eq:Omega_phi} cannot be taken for granted for heterogeneous materials. As demonstrated in subsequent sections, without special treatment, assigning a lower fracture energy to the interface merely results in an effective interfacial fracture energy that equals the average of the adjacent bulk materials. To resolve this issue, we will introduce an additional source term into the governing equations, as detailed in Section~\ref{sec:source}. This modification allows for the precise reproduction of any specified interfacial fracture energy, as well as the cohesive law defined in Eq.~\eqref{eq:Omega_phi}, as demonstrated in Section~\ref{sec:cohesive law}.
\end{remark}

\subsection{The governing equations with an interfacial source term}\label{sec:source}
Without special treatment, the diffuse nature of the phase field causes the interfacial fracture energy to be dominated by the adjacent bulk materials, resulting in values significantly higher than expected for a weak interface. To address this, we propose introducing a strongly localized compensatory phase-field source at the interface, $ q_\phi $, leveraging the strong damage localization characteristic of the $\Omega^2$ model to precisely regulate the interfacial fracture parameters.

\begin{figure}[!htbp] 
	\centering
	\includegraphics[width=0.5\linewidth]{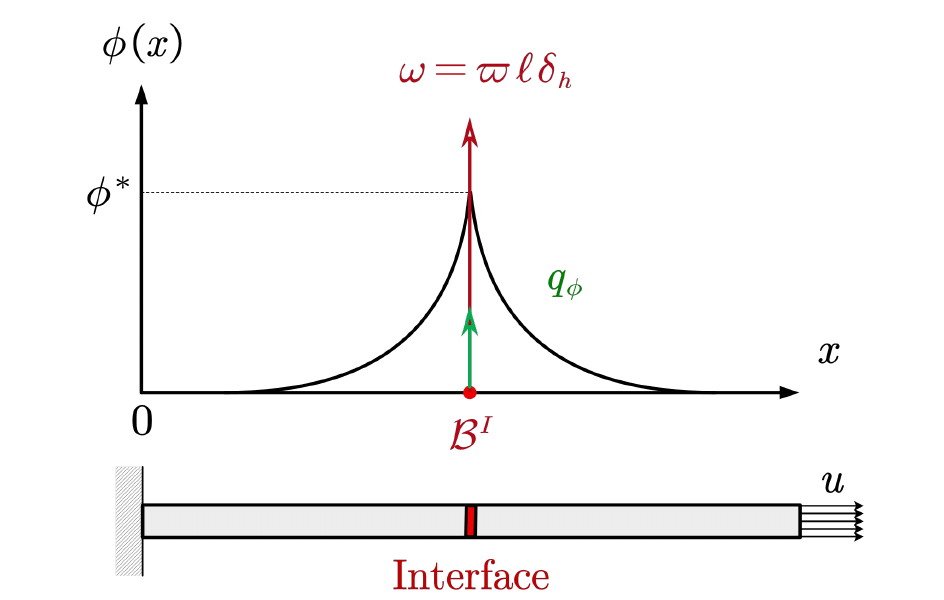}
	\caption{The localized phase-field driving source $ q_{\phi} $}
	\label{fig:yuan}
\end{figure}

In the presence of the additional source term $ q_\phi $, the first-order stability condition in traditional phase-field fracture modeling~\citep{pham2011} is extended to\footnote{Similar formulations with an additional source term can be found in \cite{feng2022phase_dissipation} and \cite{feng2024phase}.}
\begin{equation}\label{eq:deltaQ}
	\delta\mathcal{E}_\ell + \int_{\mathcal{B}} q_\phi \delta\phi \geq 0 \quad \text{for all physically feasible variations } \{\delta u, \delta\phi, \delta\omega\}.
\end{equation}
Following standard variational procedures, and utilizing integration by parts alongside the divergence theorem, the complete set of governing equations and inequalities for the model can be derived from the preceding variational conditions as follows:
\begin{align}
	& \pd{\sigma}{x} = 0;\label{eq:sigma balance}\\[0.65 em]
	&\left(\frac{\delta\mathcal{E}_\ell}{\delta\phi}+q_\phi\right)\dt{\phi}=0,\quad \frac{\delta\mathcal{E}_\ell}{\delta\phi} +  q_\phi\geq 0, \quad \text{and}\quad \dt{\phi}\geq 0;\label{eq:KT_phi}\\[0.6 em]
	&\frac{\partial\psi}{\partial\omega}\,\dt{\omega}=0,\quad \frac{\partial\psi}{\partial\omega}\geq0, \quad \text{and}\quad \dt{\omega}\geq 0.\label{eq:KT_omega}
\end{align}
 Here, the Cauchy stress $\sigma$ is derived as
 \begin{equation}
 	\sigma =\pd{\psi_{\mathrm{elastic}}}{\varepsilon}=  {E\varepsilon} -(1-g_1(\omega)){\left\langle E{\varepsilon} \right\rangle },
 \end{equation}
the functional derivative, $ \delta\mathcal{E}_\ell/\delta \phi $, is defined by $ \delta\mathcal{E}_\ell/\delta \phi =\partial_\phi\psi - \partial_x(\partial_{\partial\phi} \psi) $ and explicitly given as
\begin{equation}\label{eq:phi governing evolution}
	\frac{\delta\mathcal{E}_\ell}{\delta\phi} = \frac{G_\mathrm{c}^{j}}{\ell}\left[\Omega(\phi)\Omega(\phi)^\prime-2(1-\phi)\omega\right] -\pd{\xi}{x},\quad j\in\{A,B,I\},
\end{equation}
 the ``microscopic'' stress $\xi$ is derived as
  \begin{equation}
 	\xi = \pd{\psi_{\mathrm{crack}}}{(\partial_x\phi)}= G^j_\mathrm{c}\ell \partial_x\phi,\quad j\in\{A,B,I\},
 \end{equation}
 and, with $ \psi=\psi_{\mathrm{crack}}+\psi_{\mathrm{elastic}} $,
  \begin{equation}\label{eq:omega governing evolution}
 	\pd{\psi}{\omega}=\frac{G_\mathrm{c}^{j}}{\ell}\left( 1-\phi \right)^2+g_1^\prime(\omega)\frac{1}{2}E^j\varepsilon^2 ,\quad j\in\{A,B,I\}.
 \end{equation}
Moreover, the overdot notation $\dot{(\cdot)}$ denotes differentiation with respect to pseudo-time. Finally, the additional localized driving source term $q_\phi$ in Eq.~\eqref{eq:KT_phi} is defined as 
\begin{equation}\label{eq:Q_phi}
	q_{\phi}=-\frac{2\Delta G_\mathrm{c}^{I}}{\ell}\left( 1-\phi \right) \omega \mathit{1}_{\mathcal{B}^I}(x), \quad \text{with} \quad  \Delta G_\mathrm{c}^{I} = \frac{G_\mathrm{c}^{A}+G_\mathrm{c}^{B}}{2} -G_\mathrm{c}^{I},
\end{equation}
where $\Delta G_\mathrm{c}^{I}>0$ represents the difference between the average fracture energy of the adjacent bulk materials (denoted by $G_\mathrm{c}^{A}$ and $G_\mathrm{c}^{B}$) and the target interfacial fracture energy $G_\mathrm{c}^{I}$. Furthermore, $\mathit{1}_{\mathcal{B}^I}(x)$ is the indicator function for the interfacial region $\mathcal{B}^I$, explicitly defined as
\begin{equation}\label{eq:indicator_function}
	\mathit{1}_{\mathcal{B}^I}(x) = 
	\begin{cases} 
		1, & \text{if } x \in \mathcal{B}^I, \\ 
		0, & \text{otherwise}. 
	\end{cases}
\end{equation}
This source term $ q_\phi $ is illustrated in Fig.~\ref{fig:yuan}.

Eqs.\,\eqref{eq:sigma balance}, \eqref{eq:KT_phi}, and \eqref{eq:KT_omega} represent the fundamental evolution laws for the displacement field $u$, the phase-field $\phi$, and the damage field $\omega$, respectively.

In problems involving multiphase materials, special attention must be paid to the internal boundary conditions at the interfaces $x = a$ and $x = b$ (see Sec.~\ref{sec:1D}):
\begin{equation}
	\sigma(x^-)=\sigma(x^+),\quad \xi(x^-) = \xi(x^+),\quad x=a,b,
\end{equation}
where $x^+$ and $x^-$ denote the limits approaching $ x $ from the right (positive) and left (negative) sides, respectively.

In the absence of the additional source term $ q_\phi $, the above governing laws and internal boundary conditions essentially represent the standard first-order stability conditions for the minimum potential energy problem subject to irreversibility constraints, derived via variational principles~\cite{pham2011}.

    \subsection{The cohesive law at the interface of 1D multiphase material}\label{sec:cohesive law}

We now proceed to derive the parametric equations for the interfacial cohesive law, starting with the formulation for the cohesive stress. Assuming the interface is subjected to damage loading, Eq.~\eqref{eq:KT_omega} yields
\begin{equation}\label{eq:variation omega=0}
	\frac{G_\mathrm{c}^{I}}{\ell}\left( 1-\phi \right)^2+g_1^\prime \left( \omega \right) \frac{1}{2}E^I\varepsilon^2=0 \quad \text{in } \mathcal{B}^I.
\end{equation}
Substituting the identity $g_1^\prime(\omega)=-c_1 g_1^2(\omega)$, the 1D constitutive relation under loading $\sigma = g_1(\omega)E^I\varepsilon$, and the definition of parameter $c_1$ given in Eq.~\eqref{eq:c_1} into Eq.~\eqref{eq:variation omega=0} yields
\begin{equation}\label{eq:conhesivelaw sigma}
	\sigma^I = \left( 1-\phi^* \right) \sigma_\mathrm{c}^{I},
\end{equation}
where, invoking continuity, the phase field within the interface is assumed to take a constant value $\phi^*$ in the limit $h \to 0^+$. This result constitutes the parametric equation for the cohesive stress at the interface.

To establish the cohesive law at the interface, the displacement jump across a thin interfacial layer of width $h$ is defined as follows:
\begin{equation}\label{eq:displacement jump}
	\llbracket u \rrbracket_h := \int_a^b \varepsilon_{\mathrm{crack}} \, \mathrm{d}x = \int_a^b \left( \varepsilon - \frac{\sigma}{E} \right) \mathrm{d}x=\int_a^b c_1\omega\frac{\sigma}{E}\,\mathrm{d}x,
\end{equation}
where $\varepsilon_{\mathrm{crack}}$ denotes the crack strain defined in \cite{feng2023unified}.\footnote{An alternative formulation, $\llbracket u \rrbracket_h = \int_a^b \varepsilon \, \mathrm{d}x$, is equivalent to Eq.~\eqref{eq:displacement jump} in the limit $h \to 0^+$.}
In what follows, our target is to derive the parametric equation for $ \llbracket u \rrbracket_h $.

As proved by \cite{feng2025JMPS}, the evolution of $\omega$ in 1D problems is theoretically confined to a singular point $x^*$. It is assumed here that for $h \ll L$, the damage variable $\omega$ is non-zero exclusively at the interface.
Under this condition, the governing equation \eqref{eq:phi governing evolution} for the phase field in the adjacent bulk regions (materials A and B) reduces to
\begin{equation}\label{eq:distribution phi 1}
	\frac{G_\mathrm{c}^{j}}{\ell}\Omega(\phi)\Omega'(\phi) - \partial_x \left( G_\mathrm{c}^{j}\ell \partial_x\phi \right) = 0 \  \text{ in  }  \mathcal{B}^j, \quad  j=A,B.
\end{equation}
Since the fracture energy $G_\mathrm{c}^{j}$ is constant within each homogeneous bulk material, the gradient term simplifies to $\partial_x ( G_\mathrm{c}^{j}\ell \partial_x\phi ) = G_\mathrm{c}^{j}\ell\partial_{xx}^2 \phi$. This allows Eq.~\eqref{eq:distribution phi 1} to be unified into a single expression:
\begin{equation}\label{eq:phi_1_2}
	-\ell^2\partial_{xx}^2 \phi + \Omega(\phi)\Omega'(\phi) = 0, \quad  x \in (0,a) \cup (b,L).
\end{equation}
For a fixed time $t$, Eq.~\eqref{eq:phi_1_2} takes the form of a second-order autonomous ordinary differential equation (ODE). By employing the method of first integrals \citep{arnold1992ordinary}, this ODE can be recast as
\begin{equation}\label{eq:phi_1_2=0}
	\partial_x \big[ -\ell^2(\partial_x \phi)^2 + \Omega(\phi)^2 \big] = 0, \quad x \in (0,a) \cup (b,L).
\end{equation}
Considering the property $\Omega(0)=0$ alongside the homogeneous boundary conditions $\phi = \partial_x\phi = 0$ at the domain ends ($x=0$ and $x=L$), the integration constant vanishes, leading to
\begin{equation}\label{eq:phi_1_2==0}
	-\ell^2(\partial_x \phi)^2 + \Omega(\phi)^2 = 0, \quad x \in (0,a) \cup (b,L).
\end{equation}
Taking the square root directly yields the spatial gradient of the phase field $\phi$ in the respective bulk materials:
\begin{equation}\label{eq:distribution phi final}
	\partial_x\phi = \frac{\Omega(\phi)}{\ell}, \quad x \in (0,a) \quad \text{and} \quad \partial_x\phi = -\frac{\Omega(\phi)}{\ell}, \quad x \in (b,L).
\end{equation}

We now examine the phase-field governing equation at the interface in detail. Incorporating the source term, Eq.~\eqref{eq:phi governing evolution} yields
\begin{equation}
	\frac{G_\mathrm{c}^{I}}{\ell}\Omega(\phi)\Omega'(\phi) - \frac{2G_\mathrm{c}^{I}}{\ell}\left( 1-\phi \right)\omega - \partial_x \left( G_\mathrm{c}^{I}\ell \partial_x\phi \right) + q_\phi = 0, \quad x \in (a,b).
\end{equation}
Integrating this equation over the interfacial layer of thickness $h$ gives
\begin{equation}
	\int_a^{b=a+h} \frac{G_\mathrm{c}^{I}}{\ell}\Omega(\phi) \Omega'(\phi) \, \mathrm{d}x - \int_a^{b=a+h} \frac{2G_\mathrm{c}^{I}}{\ell}\left( 1-\phi \right)\omega \, \mathrm{d}x + \int_a^{b=a+h} q_\phi \, \mathrm{d}x = \Big[ G_\mathrm{c}^{I}\ell \partial_x\phi \Big]_a^{b=a+h}.
\end{equation}
Since the phase field $\phi$ is continuous and bounded, the first term on the left-hand side scales as $\mathcal{O}(h)$ and vanishes in the limit $h \to 0^+$. Conversely, the second term cannot be neglected because $\omega \sim 1/h$ exhibits a strong localized concentration. The right-hand side is evaluated using Eq.~\eqref{eq:distribution phi final} as $-\left( G_\mathrm{c}^{A}+G_\mathrm{c}^{B} \right)\Omega(\phi^*)$. Consequently, the equation simplifies to
\begin{equation}
	-2\frac{G_\mathrm{c}^{I}}{\ell}\left( 1-\phi^* \right)\omega h + q_\phi h = -\left( G_\mathrm{c}^{A}+G_\mathrm{c}^{B} \right)\Omega(\phi^*),
\end{equation}
where $\phi^*$ denotes the assumed uniform value of the phase field within the interface in the limit $h \to 0^+$. Substituting the definition of the source term $q_\phi$ from Eq.~\eqref{eq:Q_phi} into this relation, the expression for $\omega$ is obtained as
\begin{equation}\label{eq:omega}
	\omega = \varpi\ell\delta_h,
\end{equation}
where
\begin{equation}
	\varpi = \frac{\Omega(\phi^*)}{1-\phi^*} \quad \text{and} \quad \delta_h = \frac{1}{h}\mathit{1}_{\mathcal{B}^I}.
\end{equation}
The concise expression for $\varpi$ directly stems from the specific design of the interfacial source term $ q_\phi $.

\begin{figure}[!htbp] 
	\centering
	\includegraphics[width=0.5\linewidth]{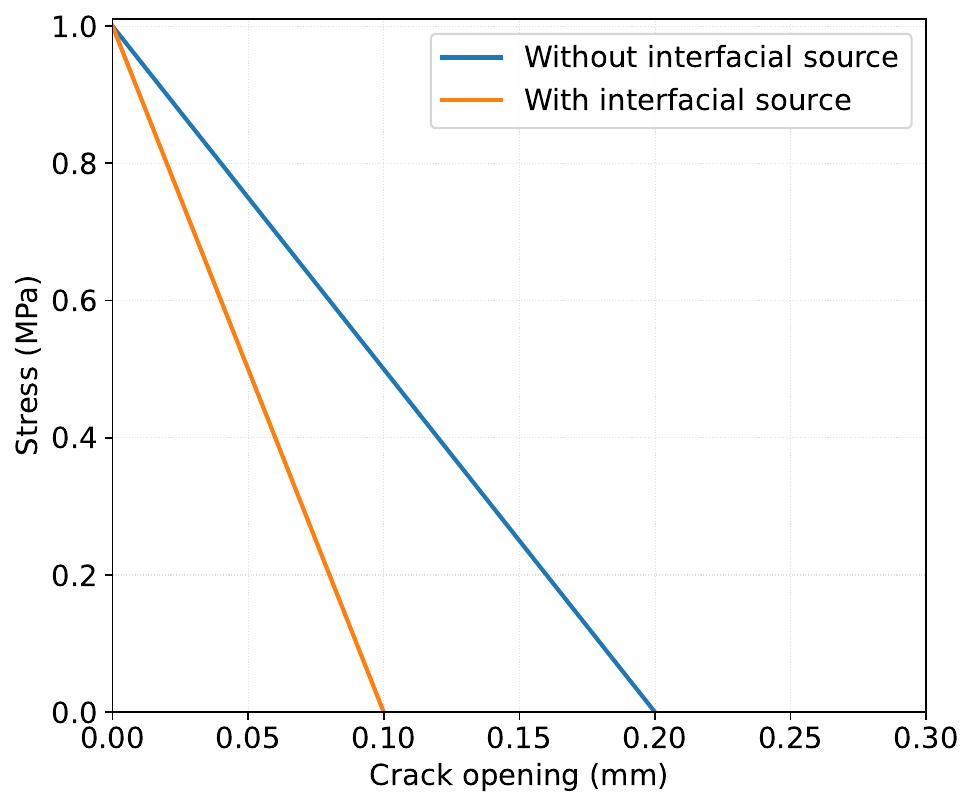}
	\caption{The cohesive laws with and without the source term: $ \Gc^I=0.1\,\mathrm{N/mm} $ and $ \bar{G}_\mathrm{c}=0.2\,\mathrm{N/mm} $.}
	\label{fig:comparison}
\end{figure}

Inserting Eq.~\eqref{eq:omega} and the parametric equation for cohesive stress, Eq.~\eqref{eq:conhesivelaw sigma}, back into the definition of the displacement jump in Eq.~\eqref{eq:displacement jump} yields
\begin{equation}
	\llbracket u \rrbracket_h = \Omega(\phi^*) \varDelta^I_{\mathrm{ch}},
\end{equation}
where $\varDelta^I_{\mathrm{ch}} = 2G_\mathrm{c}^I / \sigma_{\mathrm{c}}^I$ represents the characteristic crack opening displacement at the interface.
Taking the limit $h \to 0^+$ finally yields the parametric equations for the interfacial cohesive law:
\begin{equation}\label{eq:cohesivelaw combine h go 0}
	\sigma^I = \left( 1-\phi^* \right)\sigma_\mathrm{c}^{I} \quad \text{and} \quad \llbracket u \rrbracket = \Omega(\phi^*) \varDelta^I_{\mathrm{ch}}.
\end{equation}
If the interfacial crack opening displacement is defined as $\int_{a}^{b} \varepsilon \,\mathrm{d}x$ rather than $\int_{a}^{b} \varepsilon_\mathrm{crack} \,\mathrm{d}x$, it identically converges to Eq.~\eqref{eq:cohesivelaw combine h go 0} in the limit $h \to 0^+$. This is because the contribution from the elastic strain is of the order $\mathcal{O}(h)$, which vanishes in this limit. Furthermore, it should be emphasized that the interfacial strength $\sigma_\mathrm{c}^{I}$ in the present model is a local parameter that can be directly and accurately prescribed. Its value remains independent of whether the source term is incorporated.
\begin{remark}[Significance of the interfacial source term]
	When the source term vanishes ($q_\phi = 0$), following the same derivation procedure yields
	\begin{equation}
		\varpi = \frac{\Omega(\phi^*)}{1-\phi^*}\frac{\bar{G}_\mathrm{c}}{G_\mathrm{c}^{I}},
	\end{equation}
 with $\bar{G}_\mathrm{c} = (G_\mathrm{c}^{A} + G_\mathrm{c}^{B}) / 2$ representing the average fracture energy of the adjacent bulk materials. 
	In this case, the corresponding crack opening displacement becomes
	\begin{equation}\label{eq:cohesivelaw displacement}
		\llbracket u \rrbracket_h = \Omega(\phi^*) \bar{\varDelta}_{\mathrm{ch}},
	\end{equation}
	where $\bar{\varDelta}_{\mathrm{ch}} = 2\bar{G}_\mathrm{c} / \sigma_\mathrm{c}^{I}$. This demonstrates that omitting the source term leads to a severe overestimation of the interfacial fracture energy in phase-field simulations, posing significant risks for structural safety assessments.
Fig.~\ref{fig:comparison} illustrates this discrepancy. For example, consider a scenario where the true interfacial strength is $\sigma_{\mathrm{c}}^I=1\,\mathrm{MPa}$, the true interfacial fracture energy is $G_\mathrm{c}^I = 0.1\,\mathrm{N/mm}$, and the average fracture energy of the surrounding bulk materials is $\bar{G}_\mathrm{c} = 0.2\,\mathrm{N/mm}$. In the absence of the source term $ q_\phi $, the crack opening displacement is artificially doubled. Consequently, the area under the cohesive traction-separation curve is also doubled, leading to an erroneous overestimation of the interfacial fracture resistance.
\end{remark}

\begin{remark}\label{remark:Q_omega}
	Alongside the localized phase-field source $q_{\phi}$, a localized damage driving source $q_{\omega}$ for $ \omega $ could conceptually be incorporated:	
	\begin{equation}\label{eq:Q_omega}
		q_{\phi} = -k_\phi\frac{2 G_\mathrm{c}^{I}}{\ell}\left( 1-\phi \right) \omega \mathit{1}_{\mathcal{B}^I}(x), \quad q_{\omega} = k_{\omega} g_1^\prime\left(\omega\right)\frac{1}{2}E\varepsilon^2\mathit{1}_{\mathcal{B}^I}(x).
	\end{equation}
	Following a similar derivation procedure, one finds that the interfacial cohesive law becomes:
	\begin{equation}\label{eq:source 1+2 cohesivelaw combine h go 0}
		\sigma^I = \frac{1}{\sqrt{1+k_{\omega}}}\left( 1-\phi^* \right)\sigma_\mathrm{c}^{I} \quad \text{and} \quad \llbracket u \rrbracket_{h} = \frac{\bar{\varDelta}_{\mathrm{ch}}}{(1+k_{\phi})\sqrt{1+k_{\omega}}}\Omega \left( \phi^* \right).
	\end{equation}
	However, because material strength is a local property that can be prescribed directly, and the interfacial fracture energy can be independently modulated via $q_\phi$, introducing an additional source term for $\omega$ is deemed unnecessary. Consequently, all subsequent theoretical analyses and numerical examples in this work will exclusively consider $q_\phi$.
\end{remark}

    \section{The model in multi-dimensional space}\label{sec:3D}
	\subsection{The energy functional}
This section extends the proposed model to 2D and 3D configurations. The total potential energy of the multidimensional $\Omega^2$-model is defined as \citep{feng2025JMPS,feng_Li2026}:
\begin{equation}
	\mathcal{E}_\ell = \int_{\mathcal{B}} \Big[\psi_{\mathrm{crack}}(\phi, \nabla \phi, \omega) + \psi_{\mathrm{elastic}}(\boldsymbol{\varepsilon}, \omega)\Big] \mathrm{d}V,
\end{equation}
where the fracture surface energy density is given by
\begin{equation}
	\psi_{\mathrm{crack}} = \frac{G_{\mathrm{c}}}{2\ell} \Big[ \Omega(\phi)^2 + \ell^2 \left\| \nabla \phi \right\|^2 \Big] + \frac{G_{\mathrm{c}}}{\ell}(1-\phi)^2\omega,
\end{equation}
and the elastic strain energy density, accounting for anisotropic degradation, is expressed as
\begin{equation}\label{eq:duowei elastic}
	\psi_{\mathrm{elastic}} = \frac{1}{2} \boldsymbol{\varepsilon} : \mathbb{E} : \boldsymbol{\varepsilon} - (1-g_1(\omega)) \frac{\langle \bar{\sigma}_n \rangle^2}{2E^{\prime}} - (1-g_2(\omega)) \frac{\bar{\tau}_{n}^{2}}{2\mu}.
\end{equation}
Here, $\mathbb{E}$ denotes the isotropic stiffness tensor, and $E^\prime$ represents the effective elastic modulus, which depends on the dimensionality of the problem\footnote{Specifically, $E^\prime = E$ for 1D; $E^\prime = E/(1-\nu^2)$ for 2D plane strain; and $E^\prime = E(1-\nu)/[(1+\nu)(1-2\nu)]$ for 3D, where $\nu$ is the Poisson's ratio.}. The degradation functions for the tensile and shear modes are respectively defined as
\begin{equation}\label{eq:duowei g_1 g_2}
	g_1(\omega) = \frac{1}{1+c_1 \omega} \quad \text{and} \quad g_2(\omega) = \frac{1}{1+c_2 \omega},
\end{equation}
with the associated material constants
\begin{equation}\label{eq:duowei c_1 c_2}
	c_1 = \frac{\ell^\mathrm{ch}_1}{\ell} = \frac{G_\mathrm{c}}{\ell} \left(\frac{\sigma_{\rm c}^2}{2E^\prime}\right)^{-1} \quad \text{and} \quad c_2 = \frac{\ell^\mathrm{ch}_2}{\ell} = \frac{G_\mathrm{c}}{\ell} \left(\frac{\tau_{\rm c}^2}{2\mu}\right)^{-1},
\end{equation}
where $\sigma_{\mathrm{c}}$ and $\tau_{\mathrm{c}}$ denote the tensile and shear strengths, respectively. For the sake of simplicity, we do not explicitly distinguish material parameters other than $G_{\mathrm{c}}$ between the bulk and the interface, as these properties are assumed to be locally well-defined.

In Eq.~\eqref{eq:duowei elastic}, the nominal normal stress $\bar{\sigma}_{n}$ and shear stress $\bar{\tau}_{n}$ on the potential crack surface are
\begin{equation}\label{eq:duowei sigma_bar}
	\bar{\sigma}_n = \bar{\boldsymbol{\sigma}} : (\bmn \otimes \bmn), \quad \bar{\tau}_n = \sqrt{\| \bar{\boldsymbol{\sigma}} \cdot \bmn \|^2 - \bar{\sigma}_n^2}, \quad \text{with} \quad \bar{\boldsymbol{\sigma}} = \mathbb{E} : \boldsymbol{\varepsilon},
\end{equation}
where the crack normal $\bmn$ is determined by the strain state $\boldsymbol{\varepsilon}$ and the internal variable $\omega$ based on the principle of minimum potential energy:
\begin{equation}\label{eq:n}
	\bmn = \bmn(\boldsymbol{\varepsilon}, \omega) = \arg\min_{\|\bmn\|=1} \psi_{\mathrm{elastic}}(\boldsymbol{\varepsilon}, \omega, \bmn).
\end{equation}
The closed-form solution for the optimal orientation $\bmn$ was established for 2D in \cite{feng2023unified} and generalized to 3D in \cite{feng20253d} (see Appendix~\ref{app:crack} for details).

\subsection{Governing equations}
The first-order stability condition, incorporating the interfacial source term $q_\phi$, is expressed as:
\begin{equation}
	\delta\mathcal{E}_\ell + \int_{\mathcal{B}} q_\phi \delta\phi \, \mathrm{d}V \geq 0,
\end{equation}
which leads to the following governing equations and Karush-Kuhn-Tucker (KKT) conditions:
\begin{align}
	&\nabla \cdot \bmsi = \mathbf{0}, \label{eq:duowei EQ} \\[0.65em]	 
	&\left(\frac{\delta\mathcal{E}_\ell}{\delta\phi} + q_\phi\right) \dot{\phi} = 0, \quad \frac{\delta\mathcal{E}_\ell}{\delta\phi} + q_\phi \geq 0, \quad \dot{\phi} \geq 0, \label{eq:duowei KT_phi} \\[0.6em]
	&\frac{\partial\psi}{\partial\omega} \dot{\omega} = 0, \quad \frac{\partial\psi}{\partial\omega} \geq 0, \quad \dot{\omega} \geq 0. \label{eq:duowei KT_omega}
\end{align}

The Cauchy stress tensor $\bmsi = \partial \psi_{\mathrm{elastic}} / \partial \boldsymbol{\varepsilon}$, considering directional decomposition, is given by
\begin{equation}\label{eq:cauchy}
	\bmsi = \bar{\bmsi} - (1-g_1(\omega))\bar{\bmsi}_1^+ - (1-g_2(\omega))\bar{\bmsi}_2^+,
\end{equation}
with the directional components
\begin{align}
	\bar{\bmsi}^+_1 &= \langle\bar{\sigma}_n\rangle \bmn \otimes \bmn + \nu^\prime \langle\bar{\sigma}_n\rangle (\mathbf{I} - \bmn \otimes \bmn), \\
	\bar{\bmsi}^+_2 &= \bar{\tau}_n (\bmn \otimes \bms + \bms \otimes \bmn).
\end{align}
The actual stress components on the crack surface then satisfy
\begin{equation}\label{eq:sigma_n}
	\sigma_n = \bar{\sigma}_n - (1-g_1(\omega))\langle\bar{\sigma}_n\rangle \quad \text{and} \quad \tau_n = g_2(\omega)\bar{\tau}_n.
\end{equation}
Thus, $\sigma_n$ specializes to $g_1(\omega)\bar{\sigma}_n$ under tension ($\bar{\sigma}_n \geq 0$) and recovers $\bar{\sigma}_n$ under compression ($\bar{\sigma}_n < 0$).

The functional derivatives in Eqs.~\eqref{eq:duowei KT_phi} and \eqref{eq:duowei KT_omega} are explicitly:
\begin{align}
	\frac{\delta\mathcal{E}_\ell}{\delta\phi} &= \frac{1}{\ell} \left[ G_{\mathrm{c}}^j \Omega(\phi) \Omega'(\phi) - \ell^2 \nabla \cdot (G_{\mathrm{c}}^j \nabla \phi) \right] - \frac{2G_{\mathrm{c}}^j}{\ell} (1-\phi) \omega, \\[0.6em]
	\frac{\partial\psi}{\partial\omega} &= \frac{G_{\mathrm{c}}^j}{\ell} (1-\phi)^2 + g'_1(\omega) \frac{\langle \bar{\sigma}_n \rangle^2}{2E^\prime} + g'_2(\omega) \frac{\bar{\tau}_n^2}{2\mu},
\end{align}
where $j \in \{\mathrm{bulk, interface}\}$.

Finally, the novel source term $q_\phi$ in Eq.~\eqref{eq:duowei KT_phi} adopts a definition consistent with the 1D case:
\begin{equation}
	q_{\phi} = -\frac{2\Delta G_\mathrm{c}^{I}}{\ell}(1-\phi)\omega \mathit{1}_{\mathcal{B}^I}(\boldsymbol{x}), \quad \text{with} \quad \Delta G_\mathrm{c}^{I} = \bar{G}_\mathrm{c} - G_\mathrm{c}^{I},
\end{equation}
where $\Delta G_\mathrm{c}^{I} > 0$ represents the energy contrast between the average fracture energy of the adjacent bulk materials, $\bar{G}_\mathrm{c}$, and the target interfacial fracture energy, $G_\mathrm{c}^{I}$. The term $\mathit{1}_{\mathcal{B}^I}(\boldsymbol{x})$ denotes the indicator function for the interfacial region $\mathcal{B}^I$, as defined in Eq.~\eqref{eq:indicator_function}.
    \subsection{The unified mixed-mode cohesive law for  interfacial and bulk  fractures}
By incorporating the source term $q_\phi$, the mixed-mode (tension-shear) cohesive laws for both bulk and interfacial materials can be unified under the parametric equations proposed by Feng and Li \citep{feng2021endowing,feng2022phase,feng2023unified}:
\begin{equation}\label{eq:duowei cohesive law}
	\hat{\sigma}_{\mathrm{eff}} = 1 - \phi^* \quad \text{and} \quad \hat{\Delta}_{\mathrm{eff}} = \Omega(\phi^*),
\end{equation}
where the normalized effective stress $\hat{\sigma}_{\mathrm{eff}}$ and effective displacement jump $\hat{\Delta}_{\mathrm{eff}}$ are defined as
\begin{equation}\label{eq:eff}
	\begin{aligned}
		&\hat{\sigma}_{\mathrm{eff}} = \sqrt{\langle\hat{\sigma}_{n}\rangle^2 + \hat{\tau}_n^2}, \quad \hat{\Delta}_{\mathrm{eff}} = \sqrt{\hat{\Delta}_n^2 + \hat{\Delta}_s^2}, \\[0.5em]
		&\hat{\sigma}_n = \sigma_n / \sigma_\mathrm{c}, \quad \hat{\tau}_n = \tau_n / \tau_{\mathrm{c}}, \quad \hat{\Delta}_n = \Delta_n / \Delta_n^{\mathrm{ch}}, \quad \hat{\Delta}_s = \Delta_s / \Delta_s^\mathrm{ch}.
	\end{aligned}
\end{equation}
Here, $\Delta_n$ and $\Delta_s$ denote the normal and tangential components of the displacement jump $\llbracket \bm{u} \rrbracket$, respectively. The characteristic opening and sliding scales are given by $\Delta_n^{\mathrm{ch}} = 2G_{\mathrm{c}}/\sigma_{\mathrm{c}}$ and $\Delta_s^{\mathrm{ch}} = 2G_{\mathrm{c}}/\tau_{\mathrm{c}}$.

It is crucial to emphasize that, with the introduction of $q_\phi$, the material parameters in the aforementioned formulations (including both strength and fracture energy) strictly represent their local physical values. Consequently, the proposed model provides a consistent fracture description that seamlessly unifies the cohesive response of interfaces and bulk materials within a single mixed-mode traction--separation law.

   \subsection{The generalized $\Omega^2$-model}
   The fracture energy density can be extended to the following generalized form:
   \begin{equation}\label{eq:generalized}
   	\psi_{\mathrm{crack}} = \frac{G_{\mathrm{c}}}{2\ell} \left[ \Omega^2(\phi) + \ell^2 |\nabla \phi|^2 \right] + \frac{G_{\mathrm{c}}}{\ell} \xi(\phi)^2 \omega.
   \end{equation}
   This formulation is termed the generalized $\Omega^2$-model \citep{feng2025JMPS,feng_Li2026}. Its corresponding cohesive law is expressed in the following parametric form:
   \begin{equation}
   	\hat{\sigma}_{\mathrm{eff}} = \xi(\phi^*), \quad \hat{\Delta}_{\mathrm{eff}} = -\frac{\Omega(\phi^*)}{\xi'(\phi^*)}, \quad \phi^* \in [0, 1].
   \end{equation}
   Note that for the specific choice $\xi(\phi) = 1-\phi$, the expression recovers the original cohesive law as presented in Eq.~\eqref{eq:Omega_phi}.
   
   \subsection{A variant one-parameter model: the $p$-model}\label{sec:p}
\begin{figure}[!htbp]
	    \centering
	    \includegraphics[width=0.5\linewidth]{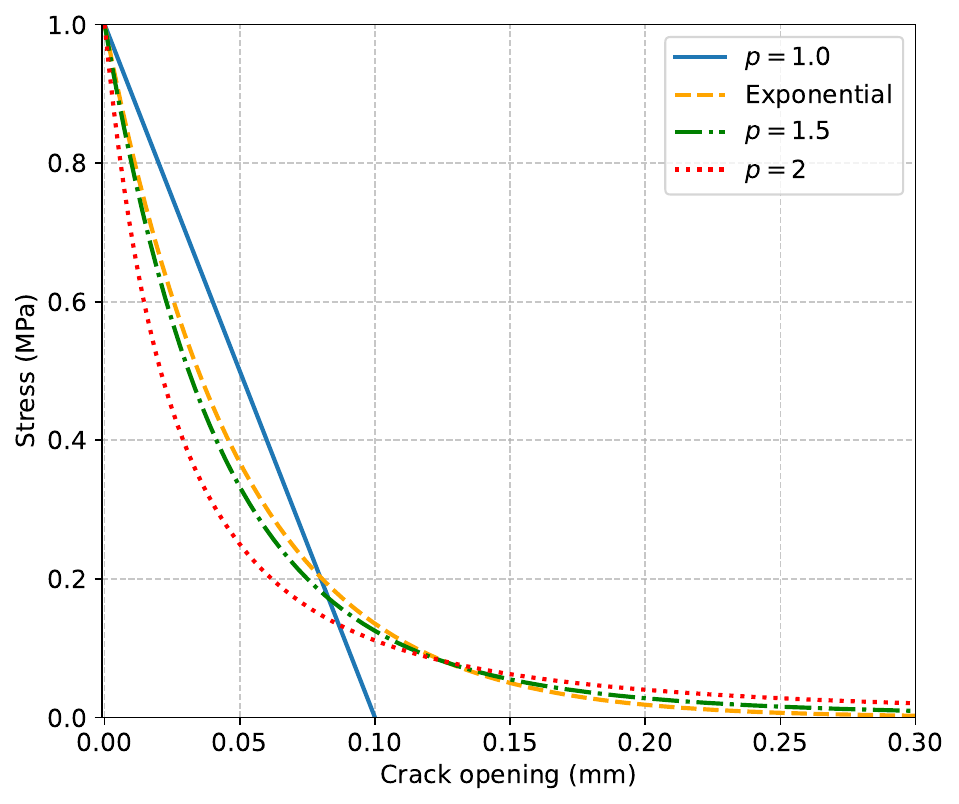}
	    \caption{Cohesive laws of exponential type and the $ p $-model for different $ p\geqslant0 $}
	    \label{fig:theoretical curve}
    \end{figure}

   By setting $\Omega(\phi) = \phi$ and $\xi(\phi) = (1-\phi)^p$ (with $p \geq 1$), we obtain a specific variant termed the $p$-model:
   \begin{equation}
   	\psi_{\mathrm{crack}} = \frac{G_{\mathrm{c}}}{2\ell} \left[ \phi^2 + \ell^2 |\nabla \phi|^2 \right] + \frac{G_{\mathrm{c}}}{\ell} (1-\phi)^{2p} \omega.
   \end{equation}
   The resulting cohesive law is given by:
   \begin{equation}
   	\hat{\sigma}_{\mathrm{eff}} = (1-\phi^*)^p, \quad \hat{\Delta}_{\mathrm{eff}} = \frac{1}{p} \phi^* (1-\phi^*)^{1-p}, \quad \phi^* \in [0, 1].
   \end{equation}
   The constitutive curves for various values of $p$ are illustrated in Fig.~\ref{fig:theoretical curve}. This single-parameter model is capable of representing a broad class of convex cohesive laws. Notably, the case $p=1$ recovers the classic linear softening law \citep{camacho1996computational}. 
   
   From a computational perspective, the $p$-model is particularly well-suited for simulating the failure of brittle and quasi-brittle materials. Due to its superior convergence rates observed in numerical experiments, this model is frequently employed in the subsequent numerical examples.

    \section{Fracture simulations}\label{sec:fracture simulation}
    
The numerical implementation in this work employs the standard staggered (alternating) iterative scheme with irreversibility constraints, the details of which are omitted here for brevity. For the local update of the damage variable $\omega$ at the Gauss points and the update of the crack normal field $\bmn$, the reader is referred to \cite{feng2025JMPS} for a comprehensive algorithmic description.
	\subsection{Example 1: Uniaxial tension of a three-phase material}\label{sec:danzhou}
\begin{figure}[!htbp] 
	\centering
	\includegraphics[width=0.7\linewidth]{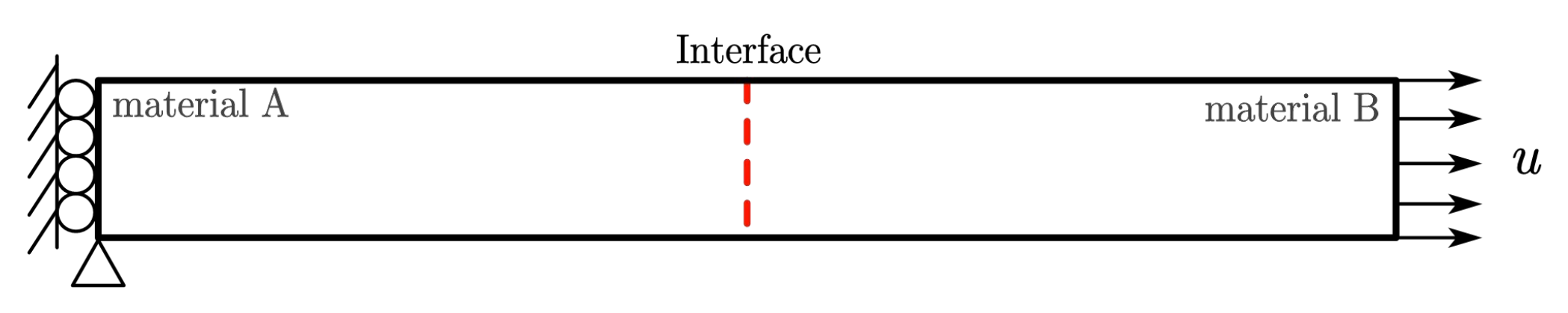}
\caption{Geometry and boundary conditions for the uniaxial tension of a three-phase material (material $ A $, material $ B $, and interface).}
	\label{fig:danzhou-gongkuang}
\end{figure}

\begin{figure}[!htbp]
	\centering
	\begin{subfigure}{.4\textwidth}
		\centering	\includegraphics[width=\linewidth]{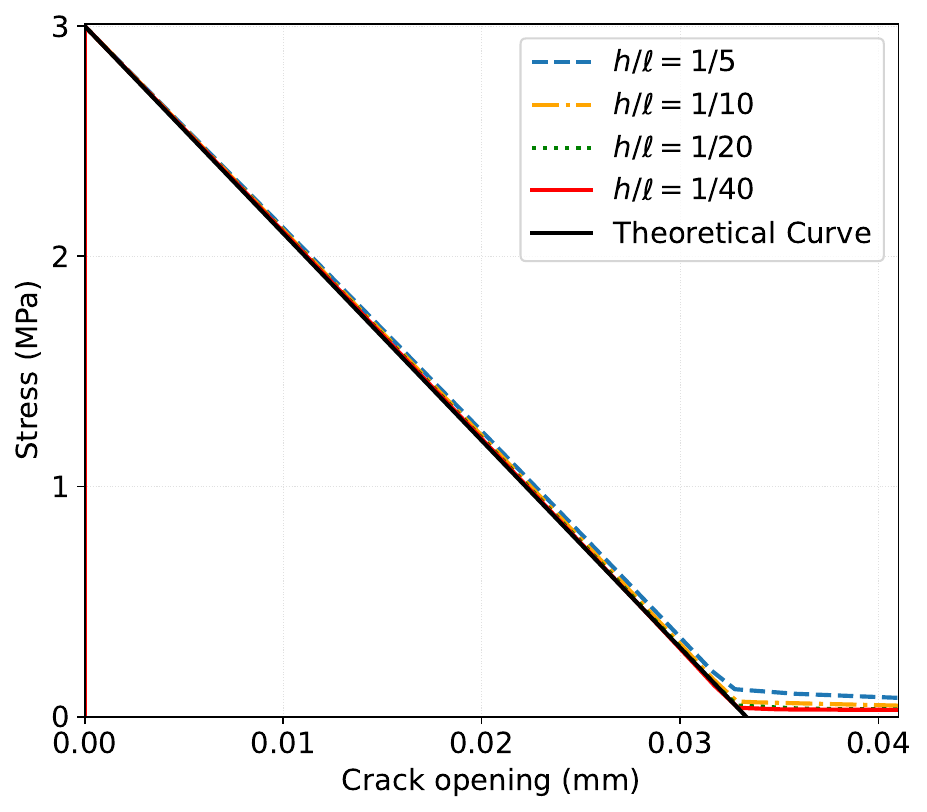}
		\caption{The linear cohesive law}
	\end{subfigure}
	\begin{subfigure}{.4\textwidth}
		\centering	\includegraphics[width=\linewidth]{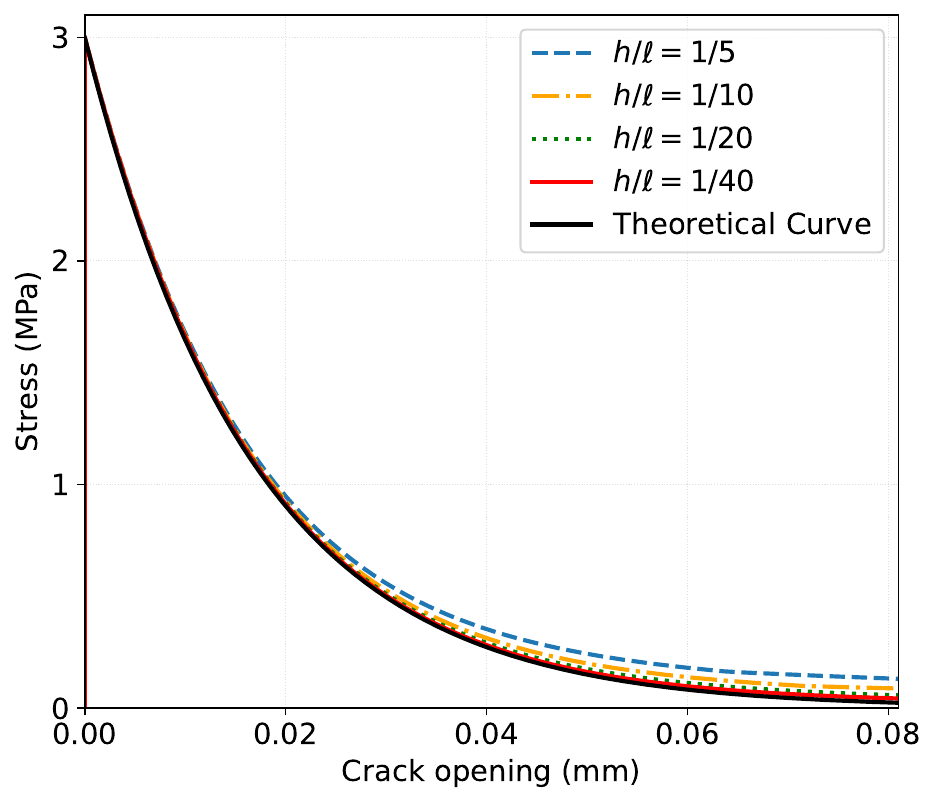}
		\caption{The exponential cohesive law}
	\end{subfigure}
		
	\vspace{0.1cm}
		
	\begin{subfigure}{.4\textwidth}
		\centering	\includegraphics[width=\linewidth]{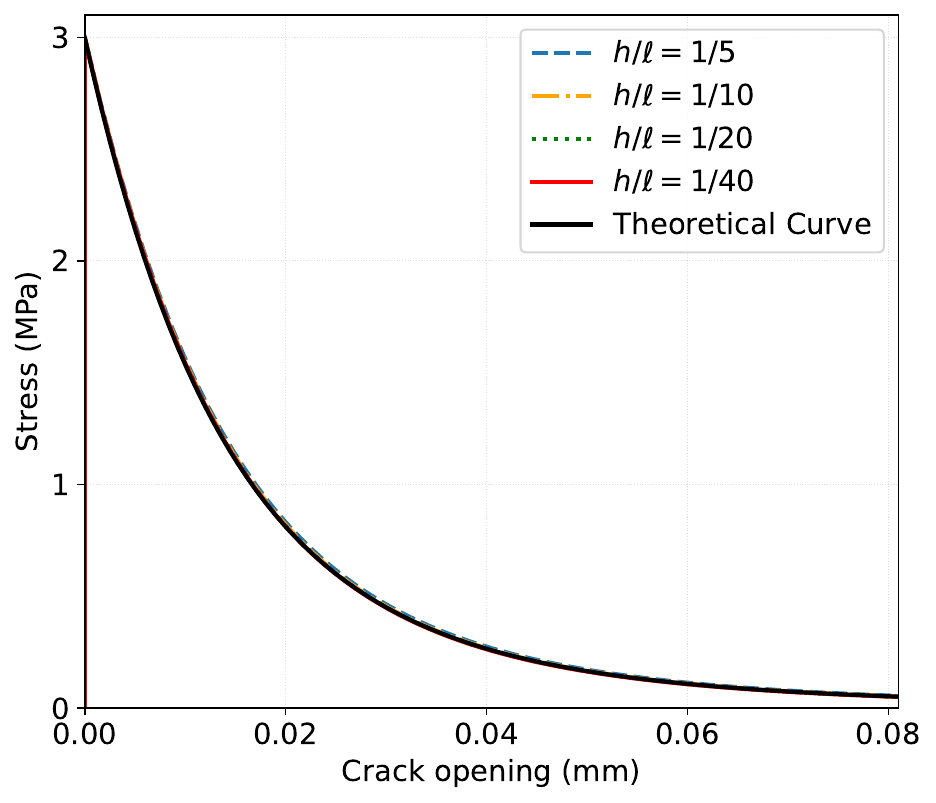}
		\caption{The cohesive law ($p=1.5$)}
	\end{subfigure}
	\begin{subfigure}{.4\textwidth}
		\centering	\includegraphics[width=\linewidth]{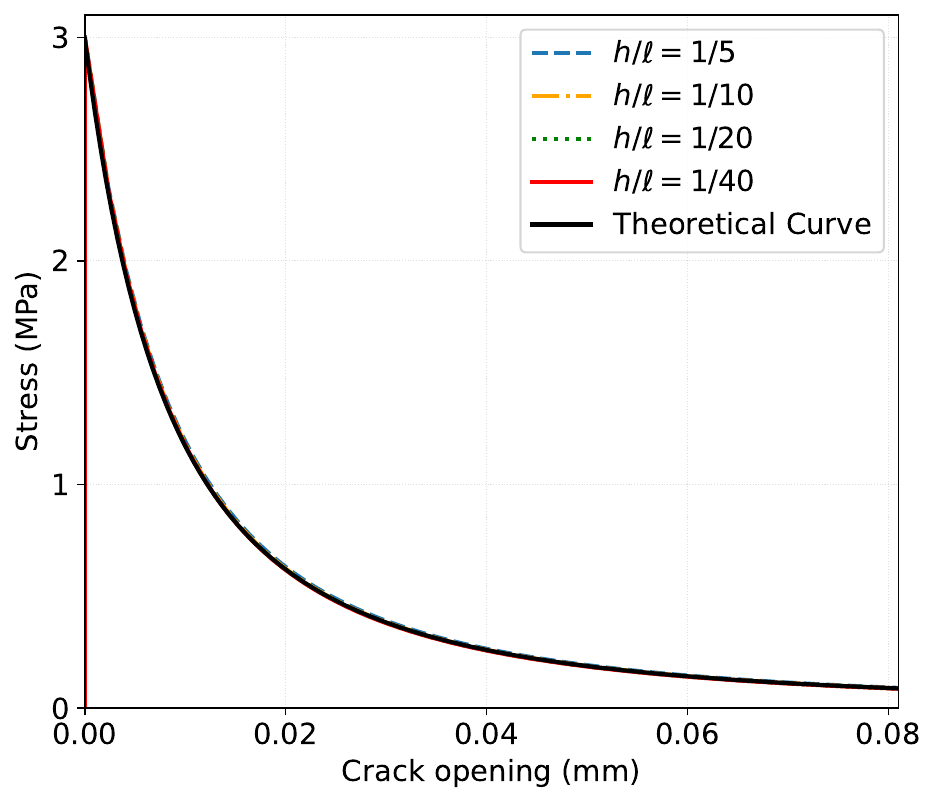}
		\caption{The cohesive law ($p=2$)}
	\end{subfigure}
	\centering
	\caption{Convergence of the model's cohesive law with respect to mesh size: simulations performed with a characteristic length $\ell=5$~mm}
	\label{fig:danzhou-cohesivelaw}
\end{figure}

In this section, we first conduct numerical simulations of the uniaxial tension test for a 2D three-phase bar---consisting of material $A$, material $B$, and an embedded interfacial layer---the theoretical basis of which was detailed in Sec.~\ref{sec:1D}. The primary objective is to verify whether the proposed model can accurately capture the constitutive responses of various established cohesive laws within heterogeneous media, while simultaneously evaluating its convergence characteristics. A secondary goal is to investigate whether the internal variable $\omega$ maintains its characteristic strong localization property in the presence of material heterogeneities.

The 2D specimen,   defined over the domain $\mathcal{B} = [0, L] \times [0, H]$ with $L = 100$~mm and $ H=10 $~mm, is investigated. The bar is clamped at the left end ($x = 0$) and subjected to a monotonically increasing longitudinal displacement at $x = L$. The specific loading configuration and boundary conditions are schematically illustrated in Fig.~\ref{fig:danzhou-gongkuang}.

The 2D bar consists of three distinct regions: material $A$ on the left, material $B$ on the right, and a thin interfacial layer situated between them. To isolate the fracture response from elastic heterogeneity, the elastic properties are assumed to be uniform across all phases, with Young's modulus $E = 5 \times 10^4$~MPa and Poisson's ratio $\nu = 0.4$. The fracture properties for each constituent are prescribed as follows:
\begin{itemize}
	\item \textbf{Material A (left):} Critical energy release rate $G_{\mathrm{c}}^A = 0.08$~N/mm and tensile strength $\sigma_{\mathrm{c}}^A = 3.795$~MPa.
	\item \textbf{Material B (right):} $G_{\mathrm{c}}^B = 0.12$~N/mm and $\sigma_{\mathrm{c}}^B = 4.648$~MPa.
	\item \textbf{Interface  (middle):} $G_{\mathrm{c}}^I = 0.05$~N/mm and $\sigma_{\mathrm{c}}^I = 3$~MPa.
\end{itemize}

\begin{figure}[!htbp] 
	\centering
	\includegraphics[width=0.85\linewidth]{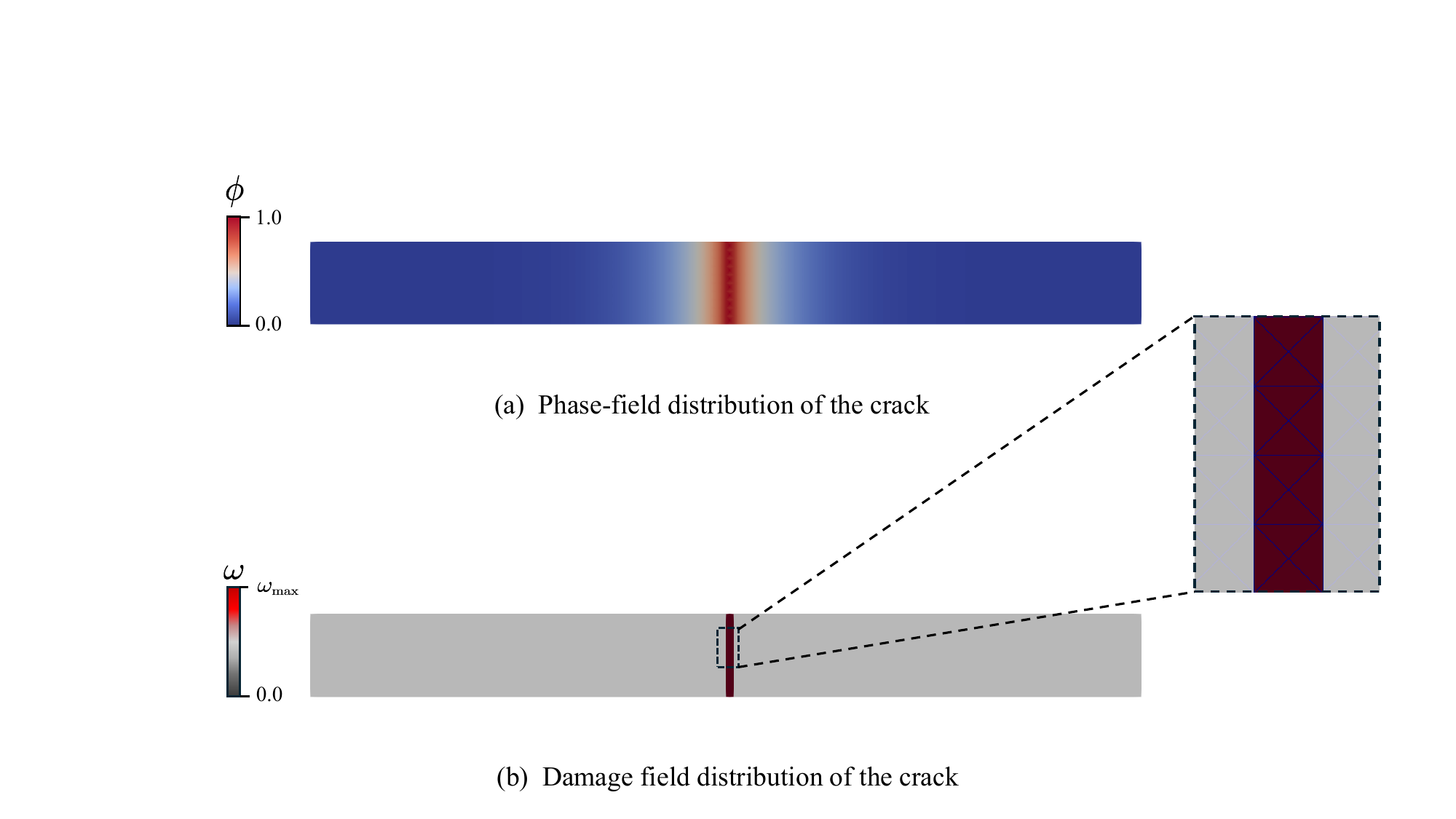}
	\caption{Comparative characterisation of crack propagation in the phase-field and damage fields under uniaxial tension}
	\label{fig:danzhou-result}
\end{figure}

Fig.~\ref{fig:danzhou-result} illustrates the evolution of the phase-field $\phi$ and the damage variable $\omega$ throughout the tensile process. As expected, fracture initiates at the interface, which represents the region of minimum tensile strength and fracture energy. 

Notably, the damage field $\omega$ exhibits strong localization at the interface, where its spatial distribution is governed by the local mesh size rather than the intrinsic bandwidth of the phase-field $\phi$. This numerical observation provides a sound verification of the fundamental assumption introduced in Sec.~\ref{sec:cohesive law}, confirming that damage evolution can be effectively decoupled from the regularization length $\ell$. Consequently, the proposed model allows for fracture to be captured within a narrow interfacial region through localized mesh refinement alone, thereby significantly enhancing computational efficiency without compromising the accuracy of the cohesive response.

The reaction force is measured at the right boundary, while the crack opening displacement (COD) is evaluated based on Eq.~\eqref{eq:displacement jump}. Numerical simulations are performed using the exponential cohesive law and the $p$-model with $p=1, 1.5, \text{and } 2$ (where $p=1$ recovers the linear softening law). The extracted numerical cohesive responses are compared with their theoretical counterparts in Fig.~\ref{fig:danzhou-cohesivelaw}.

The results demonstrate that, with successive mesh refinement, all numerical cohesive laws converge to the prescribed theoretical curves. This confirms the capability of the proposed model to accurately reproduce  cohesive constitutive relations within the interface. Notably, the $p$-model ($p = 1.5, 2$) exhibits superior convergence rates compared to both the linear ($p=1$) and exponential cohesive laws. Based on these findings, the $p$-model, characterized by its robust convergence, is adopted for all subsequent numerical investigations.

    \subsection{Example 2: Double-cantilever beam test}
The subsequent numerical example considers the Double Cantilever Beam (DCB) test, as previously studied by \citep{nguyen2016phase} and \citep{verhoosel2009dissipation}. The objective is to assess the proposed model's capability in predicting pure Mode-I interfacial fracture in 2D. This problem features crack propagation along a predefined interface, making it a classic benchmark for evaluating interfacial fracture models.

\begin{figure}[!htbp] 
	\centering
	\includegraphics[width=0.7\linewidth]{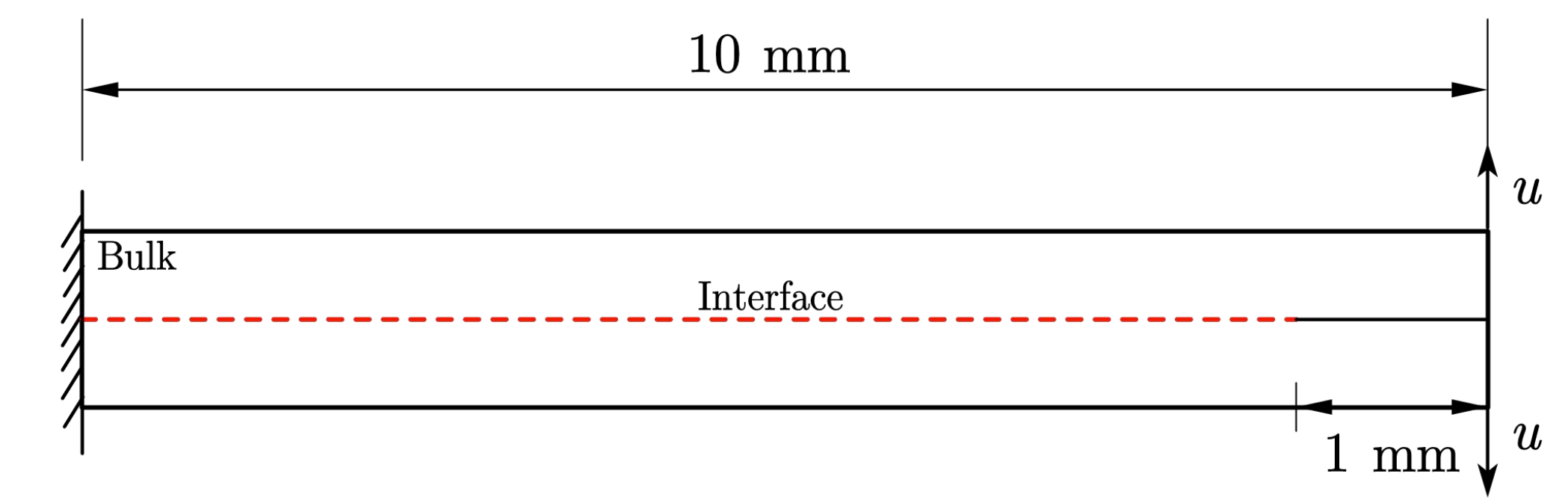}
	\caption{Schematic diagram of a double-cantilever beam specimen}
	\label{fig:DCB-gongkuang}
\end{figure}

The specimen geometry and boundary conditions are depicted in Fig.~\ref{fig:DCB-gongkuang}. The left end is fully constrained, while symmetric vertical displacements $u$ are applied to the upper and lower flanges at the right end. The beam has a length of 10~mm and a height of 1~mm. An interface is situated at the mid-plane of the beam, containing a pre-existing crack of 1~mm in length at the right end.

The upper and lower sections of the beam consist of the same bulk material, while the interface layer is embedded between them. The elastic properties for the bulk material are: Young's modulus $E = 1 \times 10^2$~MPa and Poisson's ratio $\nu = 0.3$. Its fracture properties are defined by $G_{\mathrm{c}} = 0.5$~N/mm and $\sigma_{\mathrm{c}} = 5$~MPa. For the interface, the parameters are: $E^I = 1 \times 10^3$~MPa, $\nu^I = 0.3$, $G_{\mathrm{c}}^I = 0.1$~N/mm, and $\sigma_{\mathrm{c}}^I = 1$~MPa. The interface is modeled with a thickness of $h = 0.01$~mm, corresponding to a single layer of finite elements, while the regularization length scale is prescribed as $\ell = 0.05$~mm.
In this simulation, the $p$-model with $p=1.5$ is employed. 

\begin{figure}[!htbp] 
	\centering
	\includegraphics[width=0.5\linewidth]{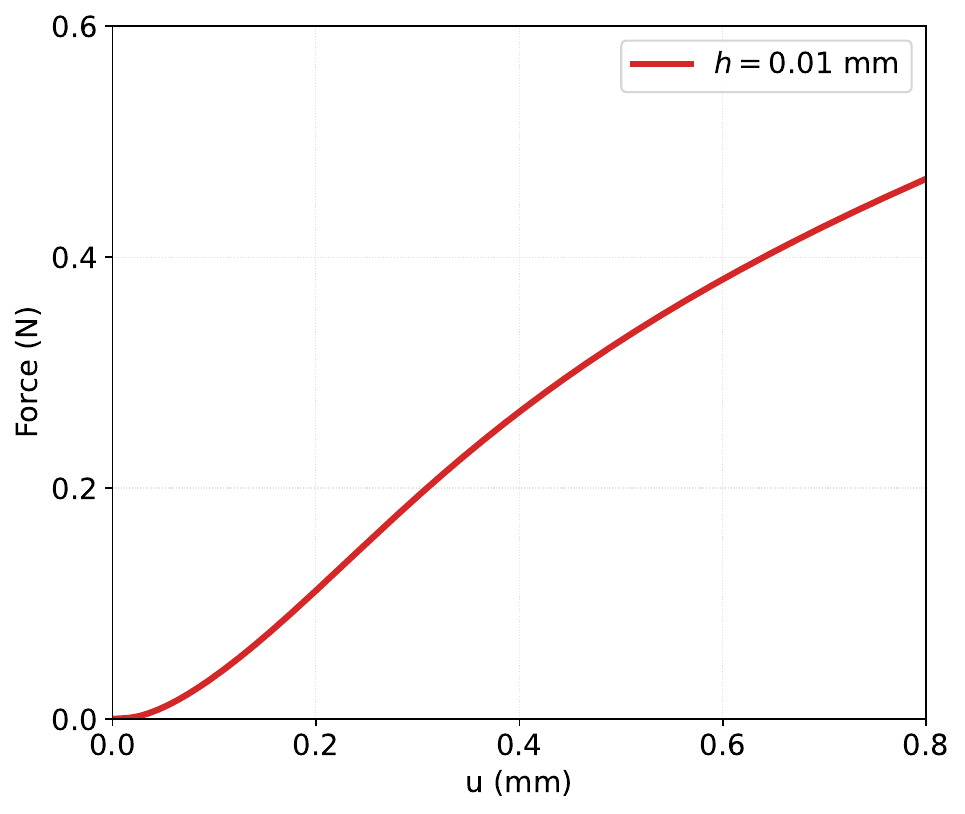}
	\caption{Force-displacement curve in the double-cantilever beam test}
	\label{fig:DCB-dis}
\end{figure}

\begin{figure}[!htbp] 
	\centering
	\includegraphics[width=0.5\linewidth]{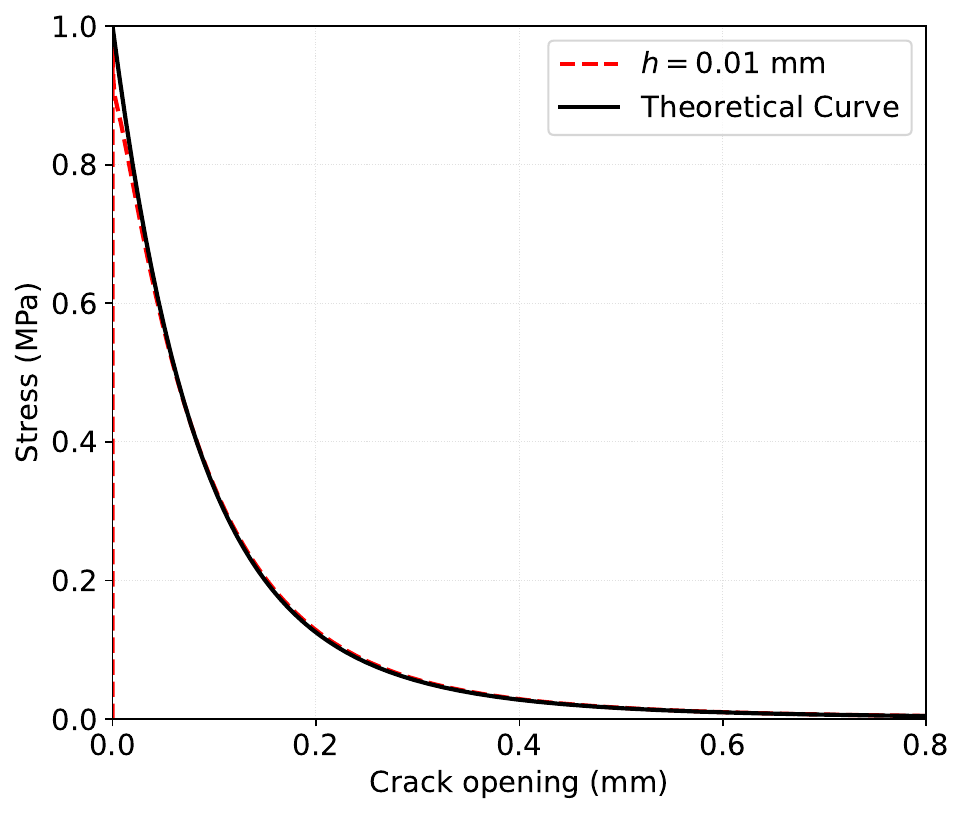}
	\caption{Two-dimensional Double-cantilever beam test: cohesive law}
	\label{fig:DCB-cohesivelaw}
\end{figure}

Regarding the mechanical response, Fig.~\ref{fig:DCB-dis} illustrates the simulated force--displacement curve, where the reaction force and corresponding displacement are measured directly at the loading points indicated in Fig.~\ref{fig:DCB-gongkuang}. Furthermore, Fig.~\ref{fig:DCB-cohesivelaw} presents the numerical cohesive law extracted using a mesh size of $h = 0.01$~mm. The crack opening displacement  is computed based on the displacement jump defined in Eq.~\eqref{eq:displacement jump}. Comparisons with the analytical solution demonstrate that, even with a relatively coarse mesh ($h/\ell = 1/5$), the simulated response aligns remarkably well with the theoretical traction--separation law.

\begin{figure}[!htbp] 
	\centering
	\includegraphics[width=0.8\linewidth]{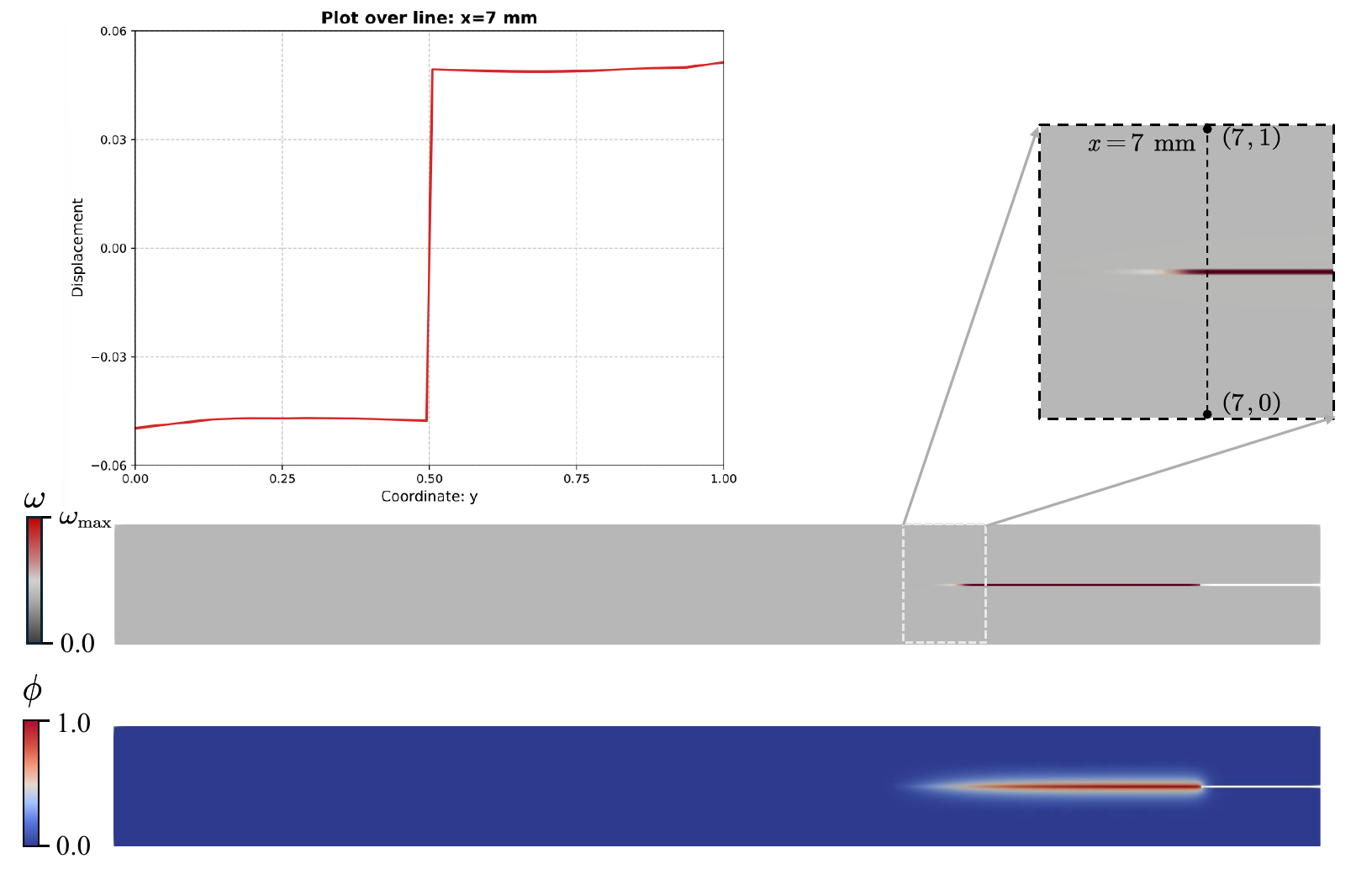}
	\caption{Comparative characterisation of crack propagation in the phase-field and damage fields in the double-cantilever beam test}
	\label{fig:DCB-result}
\end{figure}

The crack propagation results, as depicted in Fig.~\ref{fig:DCB-result}, demonstrate that the proposed model precisely predicts the fracture trajectory along the interface. \emph{It is important to emphasize that no kinematic constraints were imposed to force the crack into the interfacial region; instead, the system spontaneously selects the failure path based on the prescribed material properties and the principle of minimum potential energy.} 

Notably, the damage field $\omega$ remains strictly confined to the thin interfacial layer in this 2D configuration, exhibiting the same robust localization features observed in the 1D scenarios.
Consistent with the findings reported in \cite{feng2025JMPS} (cf. Fig.~11 therein), our results in Fig.~\ref{fig:DCB-result} also exhibit a pronounced discontinuity in the displacement field that manifests well before the full development of the phase-field band. This emergent feature, which allows for a sharp representation of the crack opening, is inherently unachievable within the framework of traditional phase-field models.

\subsection{Example 3: Single-fiber reinforced composite}

\begin{figure}[!htbp]
	\centering
	\includegraphics[width=0.45\linewidth]{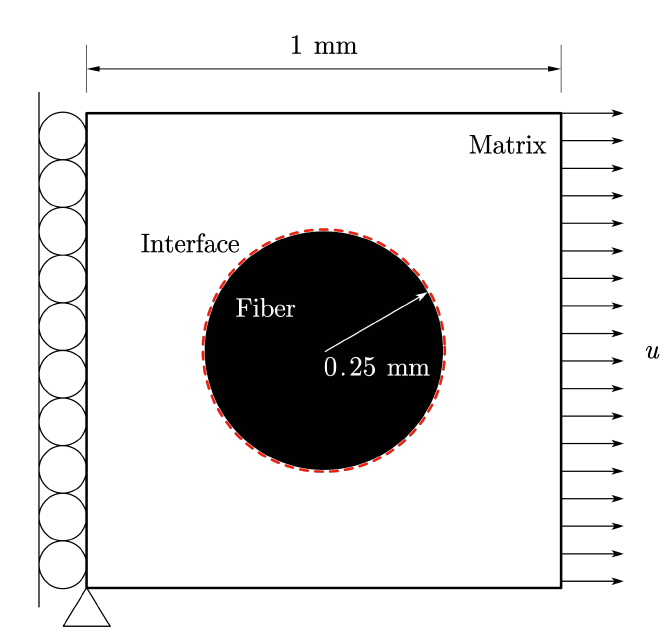}
	\caption{Problem setup for the single-fiber reinforced composite under 2D uniaxial tension: (a) geometric configuration of the matrix, fiber, and interface, and (b) prescribed boundary conditions and loading.}
	\label{fig:yi_bing-gongkuang}
\end{figure}

This example investigates a more complex failure mechanism: the fracture of a fiber-reinforced composite under tensile loading. We consider a representative volume element (RVE) containing a circular fiber, as previously studied in \citep{bian2025adaptive}. This benchmark is specifically chosen to verify the model's capability in simulating intricate crack propagation patterns governed by the competition between interfacial debonding and matrix cracking.

The geometry and boundary conditions are illustrated in Fig.~\ref{fig:yi_bing-gongkuang}. The RVE measures $1 \times 1$~mm$^2$, with a fiber of radius $R = 0.25$~mm embedded at its center. The left boundary is constrained horizontally, while the bottom-left corner is fully fixed to prevent rigid-body motion. A monotonically increasing horizontal displacement is applied to the right boundary.

The material properties for the three constituent phases are prescribed as follows:
\begin{itemize}
	\item \textbf{Matrix:} $E^M = 4 \times 10^3$~MPa, $\nu^M = 0.4$, $G_{\mathrm{c}}^M = 0.25$~N/mm, and $\sigma_{\mathrm{c}}^M = 30$~MPa.
	\item \textbf{Fiber:} $E^F = 4 \times 10^4$~MPa, $\nu^F = 0.33$, $G_{\mathrm{c}}^F = 0.5$~N/mm, and $\sigma_{\mathrm{c}}^F = 50$~MPa.
	\item \textbf{Interface:} $E^I = 3 \times 10^4$~MPa, $\nu^I = 0.4$, $G_{\mathrm{c}}^I = 0.05$~N/mm, and $\sigma_{\mathrm{c}}^I = 5$~MPa.
\end{itemize}
The interface thickness is represented by a single layer of elements with $h = 0.001$~mm, and the regularization length scale is $\ell = 0.005$~mm. The integration boundaries for evaluating the cohesive stress and the crack opening displacement (COD) are respectively defined as a small arc segment on the interface and its corresponding perpendicular cross-section, both located in the immediate vicinity of the fiber's rightmost point.

\begin{figure}[!htbp] 
	\centering
	\includegraphics[width=0.5\linewidth]{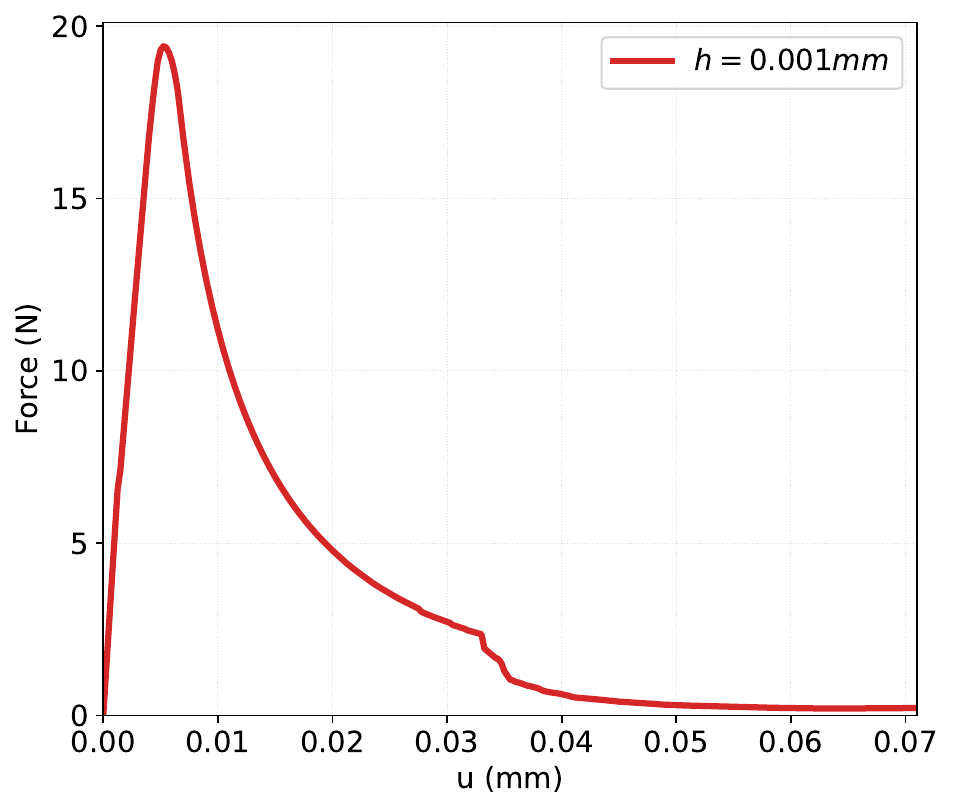}
	\caption{Simulated force-displacement curve for the single-fiber reinforced composite}
	\label{fig:yi_bing-dis}
\end{figure}

\begin{figure}[!htbp] 
	\centering
	\includegraphics[width=0.5\linewidth]{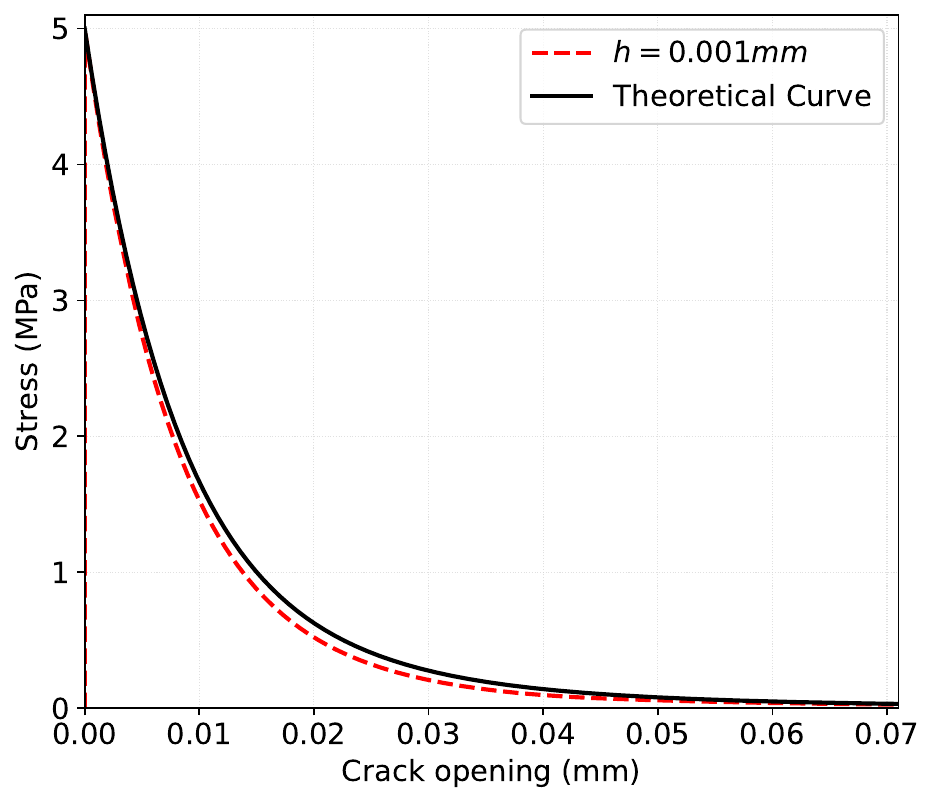}
	\caption{Comparison between the extracted numerical cohesive response at the interface and the theoretical solution for the $p$-model ($p=1.5$).}
	\label{fig:yi_bing-cohesive}
\end{figure}

Fig.~\ref{fig:yi_bing-dis} shows the predicted force-displacement curve, while Fig.~\ref{fig:yi_bing-cohesive} presents the extracted interfacial cohesive law. The numerical cohesive response shows rather good agreement with the theoretical curve. 

\begin{figure}[!htbp] 
	\centering
	\includegraphics[width=0.88\linewidth]{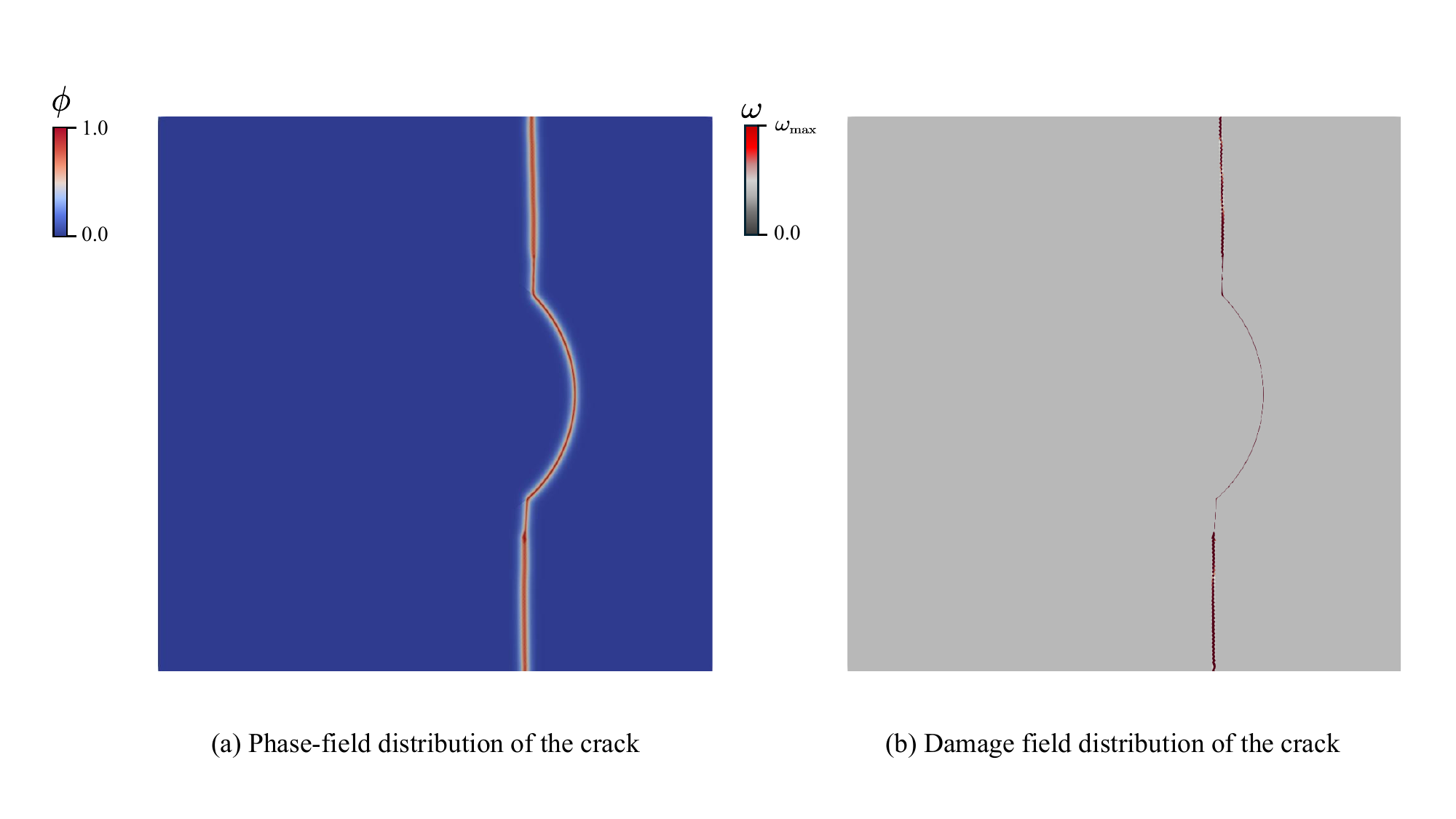} 
	\caption{Simulated failure process of the single-fiber composite: evolution of the phase-field $\phi$  and the damage field $\omega$, illustrating crack initiation, interfacial debonding, and subsequent kinking into the matrix.}
	\label{fig:yi_bing-result}
\end{figure}

As illustrated in Fig.~\ref{fig:yi_bing-result}, fracture initiates at the fiber-matrix interface, the most vulnerable region due to its significantly lower toughness. The crack preferentially propagates along the interface, leading to debonding.  Upon reaching a critical state, stress redistribution triggers a kinking instability, forcing the crack to deviate from the interface and propagate into the more resilient matrix until final failure occurs. The proposed model effectively captures this complex kinking behavior, which is intrinsically dictated by the competition between multiphase material properties and the direction-dependent fracture driving forces.

\subsection{Example 4: Crack impinging on an inclined interface}
This example evaluates the model's capacity to predict the complex interaction between a macro-crack and a heterogeneous interface---specifically, the classical competition between crack deflection (propagation along the interface) and crack penetration (crossing into the adjacent material). This benchmark, adapted from \citep{zhang2020modelling}, imposes stringent requirements on the precision of crack path predictions under mixed-mode conditions.

\begin{figure}[!htbp] 
	\centering
	\includegraphics[width=0.35\linewidth]{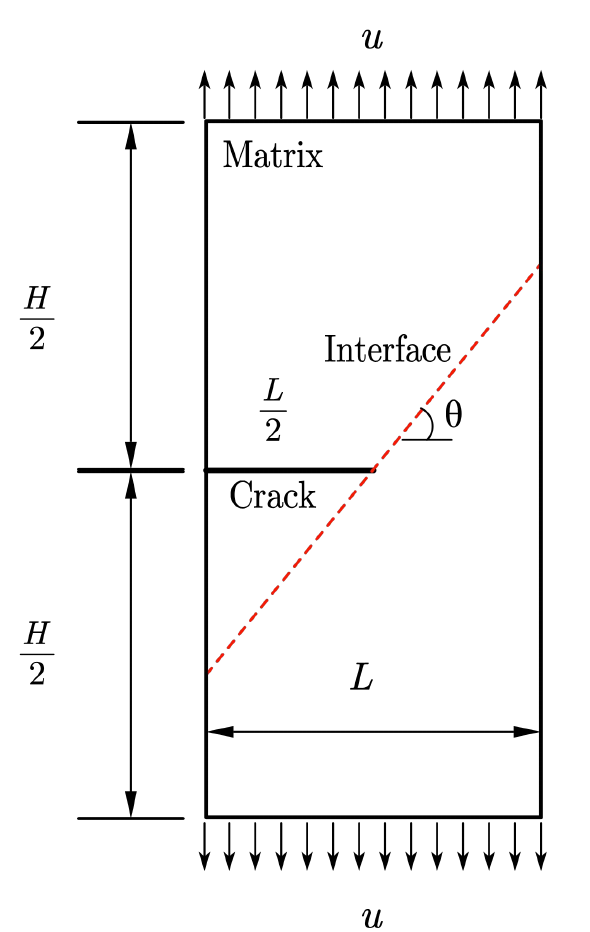}
	\caption{Geometrical configuration and boundary conditions for a plate with an inclined interface.}
	\label{fig:xie-gongkuang}
\end{figure}

The specimen consists of a rectangular plate containing a pre-existing edge crack that terminates at an inclined material interface. The detailed geometry and boundary conditions are illustrated in Fig.~\ref{fig:xie-gongkuang}. The plate dimensions are $H = 4$~mm and $L = 2$~mm, with an initial crack length of $L/2 = 1$~mm. Uniform, monotonically increasing vertical displacements are applied to the upper and lower edges. Depending on the interface properties, the failure mechanism is expected to transition between penetration and interfacial deflection.

The bulk matrix material properties are prescribed as: $E^M  = 4 \times 10^3$~MPa, $\nu^M  = 0.4$, $G_{\mathrm{c}}^M = 0.25$~N/mm, and $\sigma_{\mathrm{c}}^M  = 30$~MPa. The regularization length scale is set to $\ell = 0.02$~mm. 
 
 According to linear elastic fracture mechanics (LEFM), the competition between deflection and penetration is governed by the ratio of interface-to-matrix fracture energy $G_{\mathrm{c}}^{I}/G_{\mathrm{c}}^{M}$ and the inclination angle $\theta$. The critical condition is defined by the following relationship \citep{he1994crack,zeng2017crack,zhang2020modelling,paggi2017revisiting}:
\begin{equation}\label{eq:xie_interface-theta}
	\frac{G^{I}}{G^{M}} = \frac{1}{16} \left[ \left(3 \cos \frac{\theta}{2} + \cos \frac{3\theta}{2}\right)^2 + \left(\sin \frac{\theta}{2} + \sin \frac{3\theta}{2}\right)^2 \right]
\end{equation}
where $G^I$ and $G^{M}$ denote the energy release rates for deflection and penetration, respectively. Four parameter combinations were investigated to systematically examine this mechanism. For a ``strong" interface ($G_{\mathrm{c}}^{I}/G_{\mathrm{c}}^{M} = 0.95$), theoretical predictions suggest crack penetration for both $\theta = 30^\circ$ and $60^\circ$. Conversely, for a ``weak" interface ($G_{\mathrm{c}}^{I}/G_{\mathrm{c}}^{M} = 0.25$), deflection along the interface is anticipated.

% --- Figures ---

\begin{figure}[!htbp]
	\centering
	\begin{subfigure}{0.8\textwidth}
		\centering \includegraphics[width=\linewidth]{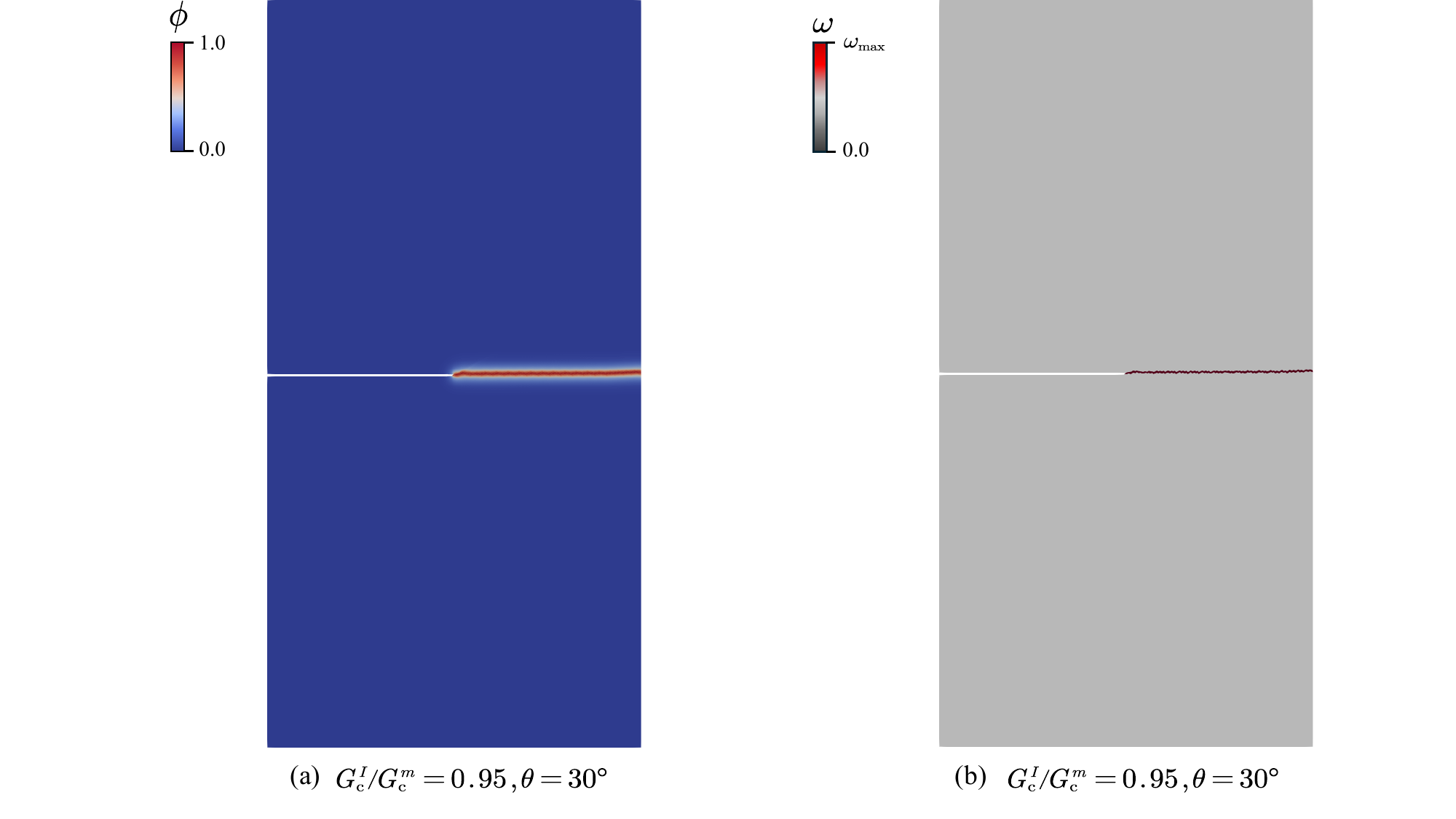}
		\caption{$G_{\mathrm{c}}^{I}/G_{\mathrm{c}}^{M} = 0.95$: Crack penetration.}
	\end{subfigure}
	\vspace{1.0em}
	
	\begin{subfigure}{0.8\textwidth}
		\centering \includegraphics[width=\linewidth]{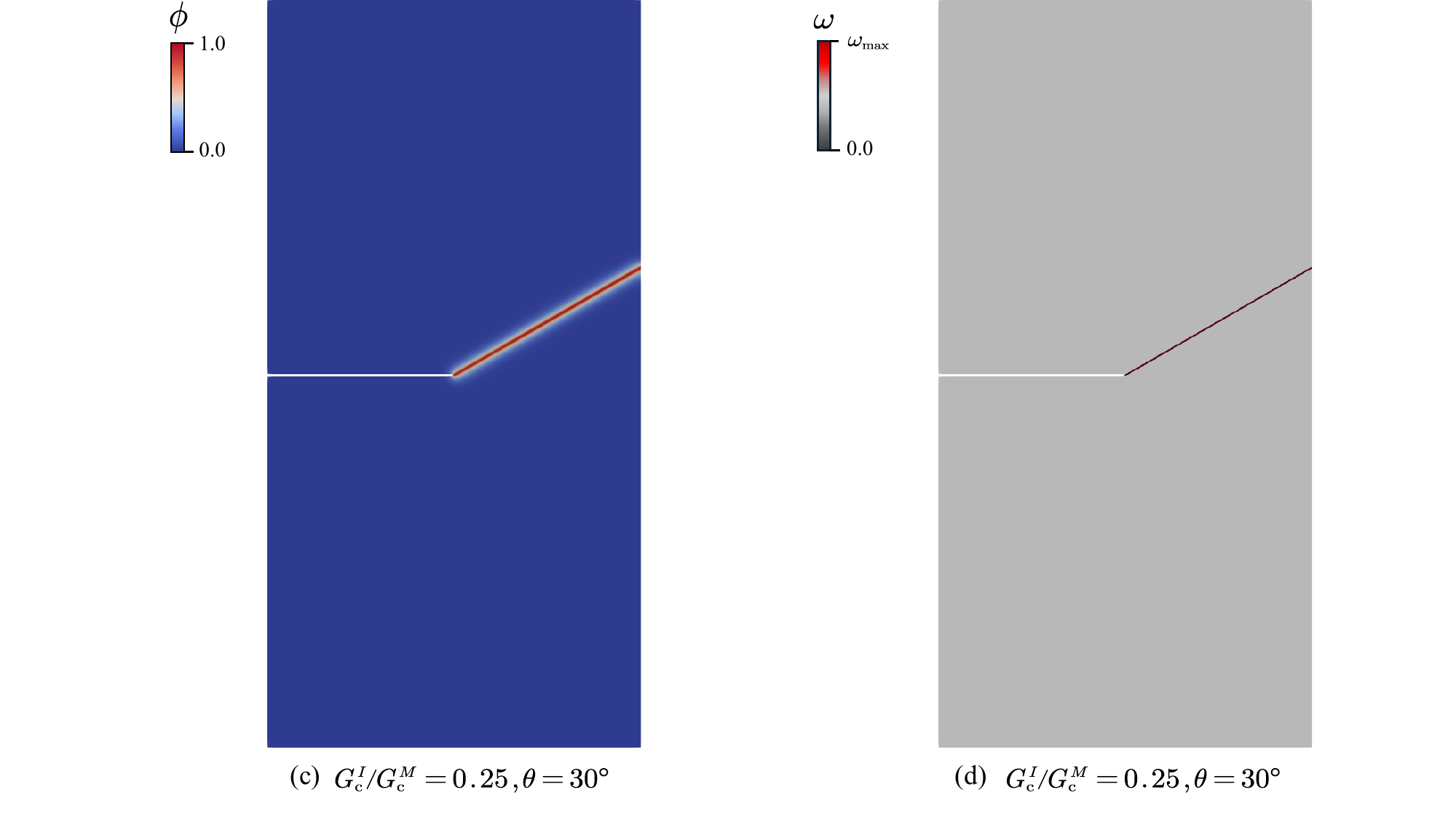}
		\caption{$G_{\mathrm{c}}^{I}/G_{\mathrm{c}}^{M} = 0.25$: Crack deflection.}
	\end{subfigure}
	\caption{Influence of interface strength on crack path selection ($\theta = 30^\circ$): Comparison between the phase-field $\phi$ (left) and the damage field $\omega$ (right).}
	\label{fig:xie-30}
\end{figure}

\begin{figure}[!htbp]
	\centering
	\begin{subfigure}{0.8\textwidth}
		\centering \includegraphics[width=\linewidth]{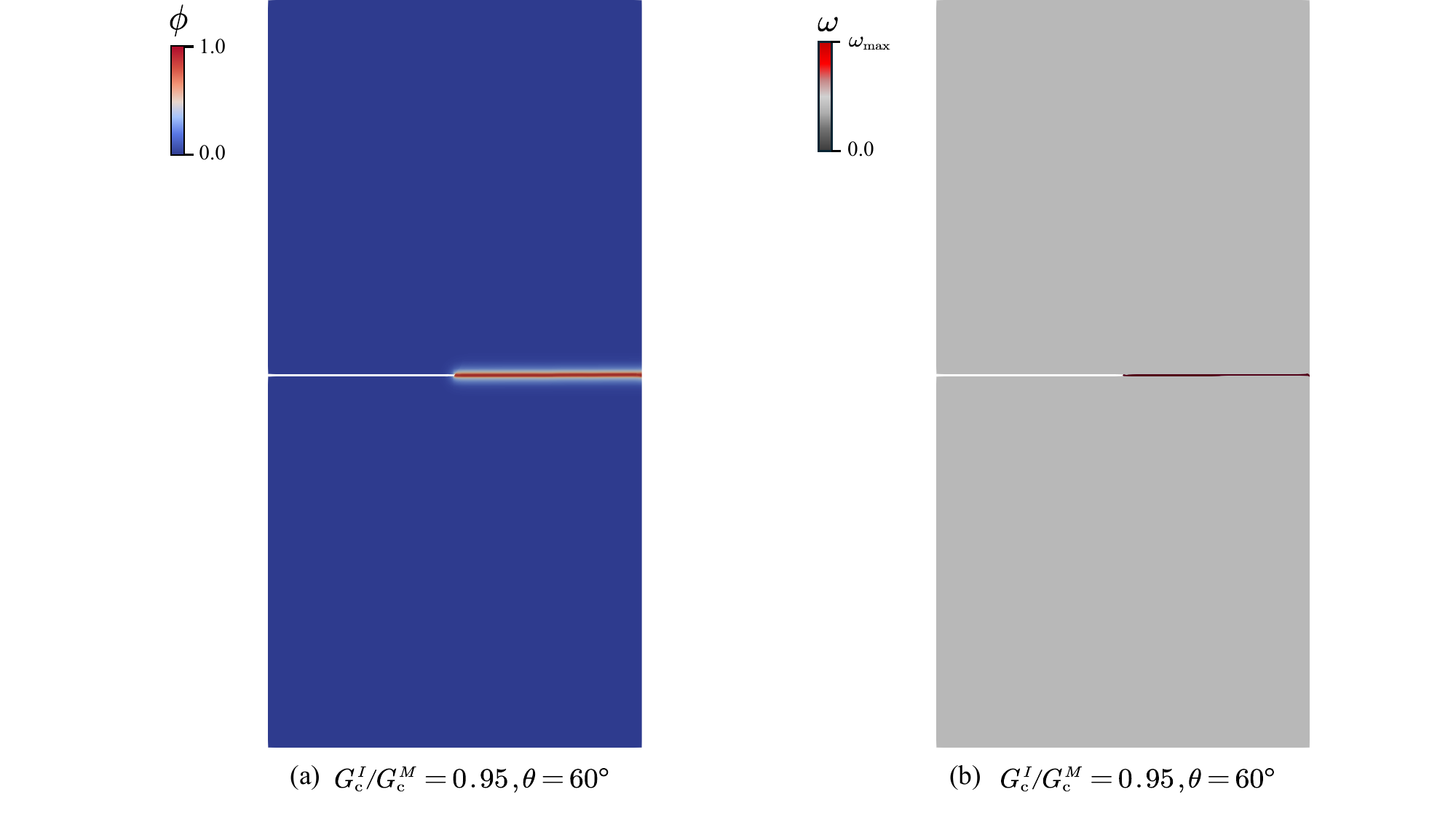}
		\caption{$G_{\mathrm{c}}^{I}/G_{\mathrm{c}}^{M} = 0.95$: Crack penetration.}
	\end{subfigure}
	\vspace{1.0em}

	\begin{subfigure}{0.8\textwidth}
		\centering \includegraphics[width=\linewidth]{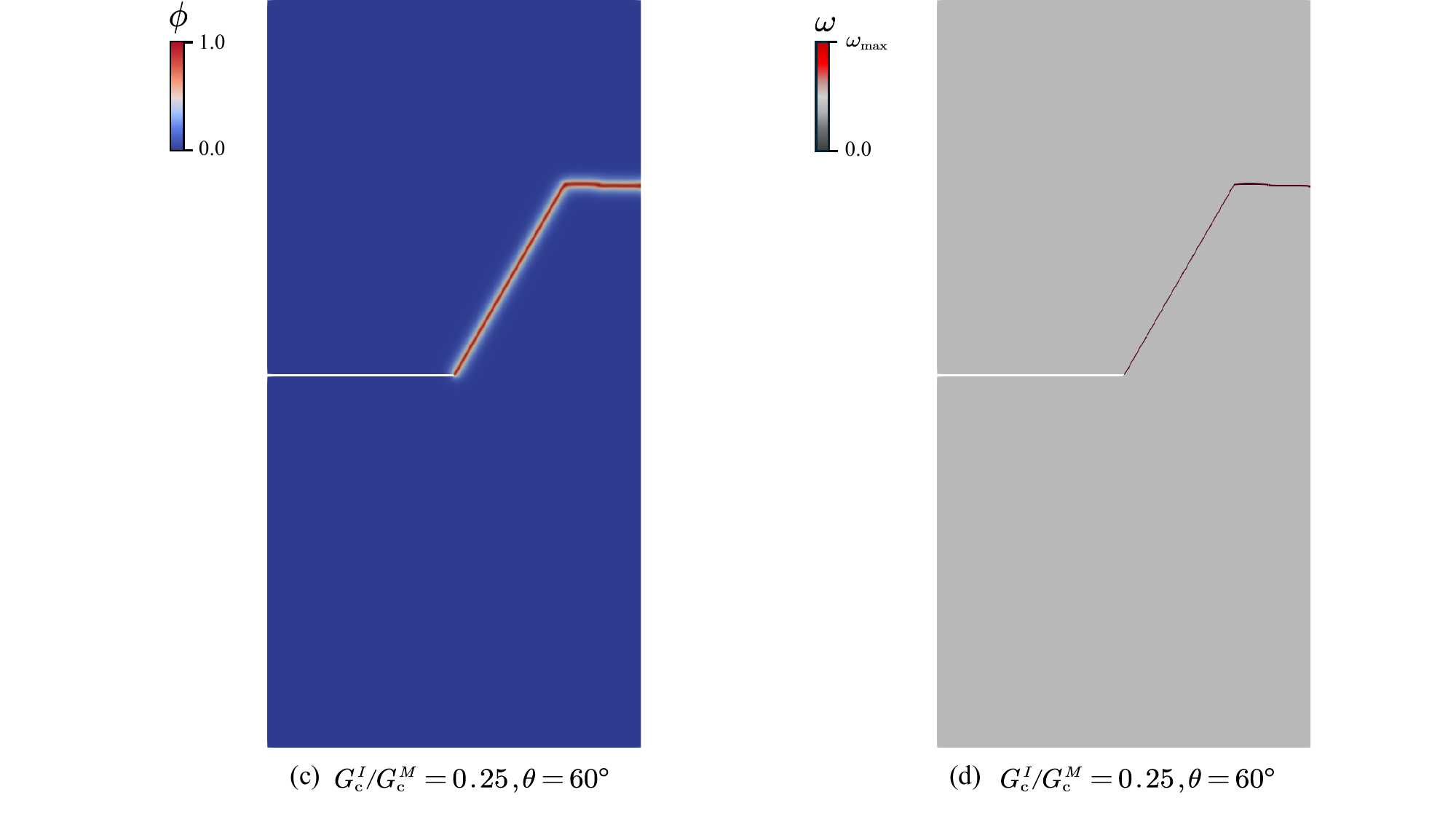}
		\caption{$G_{\mathrm{c}}^{I}/G_{\mathrm{c}}^{M} = 0.25$: Initial deflection followed by penetration.}
	\end{subfigure}
	\caption{Influence of interface strength on crack path selection ($\theta = 60^\circ$): Comparison between the phase-field $\phi$ (left) and the damage field $\omega$ (right).}
	\label{fig:xie-60}
\end{figure}
% --- End Figures ---

The simulation results are presented in Fig.~\ref{fig:xie-30} and Fig.~\ref{fig:xie-60}, showing the phase-field $\phi$ and damage field $\omega$ distributions. For $\theta = 30^\circ$, the contours clearly distinguish between penetration (strong interface) and deflection (weak interface). For $\theta = 60^\circ$, the model captures a more nuanced failure mode in the weak interface case: the crack initially deflects along the interface but subsequently becomes unstable and penetrates back into the matrix. This transitional behavior is highly consistent with theoretical expectations as the energy release rate ratio approaches the critical threshold.

These findings demonstrate that the proposed model accurately predicts the deflection of fracture paths at interfaces based on their energetic and geometric characteristics. This further validates the model's robustness in simulating complex fracture phenomena in heterogeneous media.
   
\section{Conclusion}\label{sec:conclusion}
This study develops a unified sharp-diffusive phase-field model for multiphase materials by incorporating a strongly localized interfacial source term $q_{\phi}$. A key feature of this framework is its inheritance of the $\Omega^2$ phase-field theory's capacity to manifest emergent strong discontinuities. By leveraging the strong localization property of the damage field $\omega$, the model enables the seamless representation of sharp displacement jumps within a continuum setting. The introduction of the source term $q_{\phi}$ further advances this theory, allowing for the independent and precise control of interfacial fracture energy, thereby bridging the gap between localized cohesive zone responses and diffusive phase-field representations.

Theoretical analysis and numerical benchmarks collectively demonstrate that the proposed model provides a straightforward and consistent framework for describing cohesive failure in both bulk and interfacial regions, unified by the parameterized cohesive laws of Feng and Li \citep{feng2021endowing,feng2022phase,feng2022phase_dissipation,feng2023unified}. This unified representation accurately captures the critical competition between fracture toughness and crack driving forces across heterogeneous domains, which dictates the complexity of crack path evolution. Consequently, the proposed unified sharp-diffusive phase-field model may offer a robust and computationally efficient approach for simulating intricate fracture phenomena in heterogeneous media.

Finally, we note that several recent studies~\citep{bourdin2025variational,vicentini2025variational} have also demonstrated characteristics of strong discontinuity and damage localization. The methodology proposed in this paper can be readily extended to these models.
\section*{Acknowledgments} 
Ye Feng gratefully acknowledges financial support from the National Natural Science Foundation of China (Grant No. 12302101).  This work is also funded by Key Laboratory of Design and Manufacturing Technologies for Composite Structures, Ministry of Education.
\appendix
   \section{The optimal crack orientation}\label{app:crack}
   The ``mesoscopic" elastic strain energy density is defined as \citep{feng2023unified,feng20253d,feng2025JMPS}:
   \begin{equation}\label{eq:meso}
   	\psi_{\mathrm{elastic}}^{\me}(\bme, \omega, \bmn) = \frac{1}{2} \boldsymbol{\varepsilon} : \mathbb{E} : \boldsymbol{\varepsilon} - (1-g_1(\omega)) \frac{\langle \bar{\sigma}_{n} \rangle^2}{2E'} - (1-g_2(\omega)) \frac{\bar{\tau}_n^2}{2\mu},
   \end{equation}
   where the crack normal $\bmn$ is treated as an independent kinematic variable. The minimization problem in Eq.~\eqref{eq:n} can be equivalently re-expressed as the following maximization:
   \begin{equation}\label{eq:n'}
   	\bmn = \arg\max_{\|\bmn\|=1} \Big[ \langle \bar{\sigma}_n \rangle^2 + \beta(\omega)^2 \bar{\tau}_n^2 \Big],
   \end{equation}
   where 
   \begin{equation}\label{eq:beta}
   	\beta(\omega)^2  = \frac{g_2(\omega)}{g_1(\omega)} \frac{\sigma_{\rm c}^2}{\tau_{\rm c}^2}
   \end{equation}
   represents the modified tensile-shear strength ratio, which generally differs from the intrinsic ratio $\beta_0 = \sigma_{\rm c} / \tau_{\rm c}$.
   
   Eq.~\eqref{eq:n'} admits a closed-form solution in both 2D and 3D \citep{feng2023unified,feng20253d}, given by:
   \begin{equation}\label{eq:n equation}
   	\bmn = \cos\theta \, \bar{\bm{e}}_1 + \sin\theta \, \bar{\bm{e}}_d,
   \end{equation}
   where $\bar{\bm{e}}_1$ and $\bar{\bm{e}}_d$ denote the major and minor principal directions of the undamaged stress $\bar{\bmsi} = \mathbb{E} : \bme$, respectively. The corresponding slip direction $\bms$ used in Eq.~\eqref{eq:duowei sigma_bar} is:
   \begin{equation}\label{eq:s}
   	\bms = \sin\theta \, \bar{\bm{e}}_1 - \cos\theta \, \bar{\bm{e}}_d.
   \end{equation}
   The optimal angle $\theta$ between $\bmn$ and $\bar{\bm{e}}_1$ within the $\bar{\sigma}_1$--$\bar{\sigma}_d$ plane is determined by the adjusted strength ratio $\beta(\omega)$ and the current stress state:
   \begin{itemize}
   	\item If $\beta(\omega) \leq 1$, then 
   	\begin{equation}\label{eq:theta_opt1}
   		\theta = \begin{cases}
   			0, & \text{if } \bar{\sigma}_1 \geq \beta(\omega) \bar{\tau}_{\rm max}, \\
   			\pi/4, & \text{if } \bar{\sigma}_1 < \beta(\omega) \bar{\tau}_{\rm max},
   		\end{cases}
   	\end{equation}
   	where $\bar{\tau}_{\rm max} = (\bar{\sigma}_1 - \bar{\sigma}_d)/2$.
   	\item  If $\beta(\omega) > 1$, then 
   	\begin{equation}\label{eq:theta_opt2}
   		\theta = \begin{cases}
   			0, & \text{if } \bar{\sigma}_d \geq \kappa \bar{\sigma}_1, \\
   			\frac{1}{2} \arccos \left[ \frac{1}{\beta(\omega)^2-1} \frac{\bar{\sigma}_1 + \bar{\sigma}_d}{\bar{\sigma}_1 - \bar{\sigma}_d} \right], & \text{if } \kappa \bar{\sigma}_1 > \bar{\sigma}_d > -\bar{\sigma}_1, \\
   			\pi/4, & \text{if } \bar{\sigma}_d \leq -\bar{\sigma}_1,
   		\end{cases}
   	\end{equation}
   	with $\kappa = 1 - 2/\beta(\omega)^2$.
   \end{itemize}
   
Importantly, the crack orientation $\bmn = \bmn(\bmsi)$ in the stress space is found to depend exclusively on the stress state, remaining independent of the damage variable $\omega$. This pivotal conclusion was initially established for 2D problems by \cite{feng2023unified} and subsequently generalized to 3D cases via Legendre transformation in \cite{feng20253d}. Such an orientation criterion is inherently self-consistent, as it aligns with both the maximum driving force principle and the minimum potential energy principle. From a practical standpoint, this theoretical consistency provides a robust foundation for tracking complex crack propagation paths and significantly enhances numerical stability during multi-axial loading simulations.

\bibliographystyle{elsarticle-harv}
\bibliography{ref_temp}

@article{feng2025JMPS,
title = {Phase-field cohesive fracture models with strong displacement discontinuities},
journal = {Journal of the Mechanics and Physics of Solids},
volume = {208},
pages = {106479},
year = {2025},
issn = {0022-5096},
author = {Ye Feng and Lu Hai}
}

@article{pham2011,
title={The issues of the uniqueness and the stability of the homogeneous response in uniaxial tests with gradient damage models},
author={Pham, Kim and Marigo, Jean-Jacques and Maurini, Corrado},
journal={Journal of the Mechanics and Physics of Solids},
volume={59},
number={6},
pages={1163--1190},
year={2011},
publisher={Elsevier}
}

@article{feng20253d,
  title={{3D} phase-field cohesive fracture: Unifying energy, driving force, and stress criteria for crack nucleation and propagation direction},
  author={Feng, Ye and Hai, Lu},
  journal={Journal of the Mechanics and Physics of Solids},
  pages={106036},
  year={2025},
  publisher={Elsevier}
}

@article{feng2021endowing,
  title={Endowing explicit cohesive laws to the phase-field fracture theory},
  author={Feng, Ye and Fan, Jiadi and Li, Jie},
  journal={Journal of the Mechanics and Physics of Solids},
  pages={104464},
  year={2021},
  publisher={Elsevier}
}

@article{feng2023unified,
  title={A unified regularized variational cohesive fracture theory with directional energy decomposition},
  author={Feng, Ye and Li, Jie},
  journal={International Journal of Engineering Science},
  volume={182},
  pages={103773},
  year={2023},
  publisher={Elsevier}
}

@article{feng_Li2026, 
  title={Convergence of phase-field models with emergent discontinuities to CZMs in
multi-dimension (under review)},
  author={Feng, Ye and Li, Jie},
  journal={Journal of the Mechanics and Physics of Solids},
  year={2026},
  publisher={Elsevier}
}

@article{feng2022phase,
  title={Phase-field cohesive fracture theory: A unified framework for dissipative systems based on variational inequality of virtual works},
  author={Feng, Ye and Li, Jie},
  journal={Journal of the Mechanics and Physics of Solids},
  volume={159},
  pages={104737},
  year={2022},
  publisher={Elsevier}
}

@article{feng2022phase_dissipation,
  title={Phase-field method with additional dissipation force for mixed-mode cohesive fracture},
  author={Feng, Ye and Li, Jie},
  journal={Journal of the Mechanics and Physics of Solids},
  volume={159},
  pages={104693},
  year={2022},
  publisher={Elsevier}
}

@article{feng2024phase,
  title={Phase-field model for 2D cohesive-frictional shear fracture: An energetic formulation},
  author={Feng, Ye and Freddi, Francesco and Li, Jie and Ma, Yu E},
  journal={Journal of the Mechanics and Physics of Solids},
  volume={189},
  pages={105687},
  year={2024},
  publisher={Elsevier}
}

@article{he1994crack,
  title={Crack deflection at an interface between dissimilar elastic materials: role of residual stresses},
  author={He, Ming Yuan and Evans, Anthony G and Hutchinson, John W},
  journal={International Journal of Solids and Structures},
  volume={31},
  number={24},
  pages={3443--3455},
  year={1994},
  publisher={Elsevier}
}

@article{zeng2017crack,
  title={Crack deflection in brittle media with heterogeneous interfaces and its application in shale fracking},
  author={Zeng, Xiaguang and Wei, Yujie},
  journal={Journal of the Mechanics and Physics of Solids},
  volume={101},
  pages={235--249},
  year={2017},
  publisher={Elsevier}
}

@article{bourdin2000numerical,
  title={Numerical experiments in revisited brittle fracture},
  author={Bourdin, Blaise and Francfort, Gilles A and Marigo, Jean-Jacques},
  journal={Journal of the Mechanics and Physics of Solids},
  volume={48},
  number={4},
  pages={797--826},
  year={2000},
  publisher={Elsevier}
}

@article{nguyen2016phase,
  title={A phase-field method for computational modeling of interfacial damage interacting with crack propagation in realistic microstructures obtained by microtomography},
  author={Nguyen, Thanh-Tung and Yvonnet, Julien and Zhu, Q-Z and Bornert, Michel and Chateau, Camille},
  journal={Computer Methods in Applied Mechanics and Engineering},
  volume={312},
  pages={567--595},
  year={2016},
  publisher={Elsevier}
}

@article{verhoosel2009dissipation,
  title={A dissipation-based arc-length method for robust simulation of brittle and ductile failure},
  author={Verhoosel, Clemens V and Remmers, Joris JC and Guti{\'e}rrez, Miguel A},
  journal={International journal for numerical methods in engineering},
  volume={77},
  number={9},
  pages={1290--1321},
  year={2009},
  publisher={Wiley Online Library}
}

@article{bian2025adaptive,
  title={Adaptive phase-field cohesive-zone model for simulation of mixed-mode interfacial and bulk fracture in heterogeneous materials with directional energy decomposition},
  author={Bian, Pei-Liang and Liu, Qinghui and Zhang, Heng and Qing, Hai and Schmauder, Siegfried and Yu, Tiantang},
  journal={Computer Methods in Applied Mechanics and Engineering},
  volume={443},
  pages={118062},
  year={2025},
  publisher={Elsevier}
}

@article{bian2025variationally,
  title={A variationally-consistent phase-field cohesive zone model for mixed-mode fracture with directional energy decomposition scheme and modified-G criterion},
  author={Bian, Pei-Liang and Qing, Hai and Schmauder, Siegfried and Yu, Tiantang},
  journal={International Journal of Engineering Science},
  volume={210},
  pages={104223},
  year={2025},
  publisher={Elsevier}
}

@article{zhang2020modelling,
  title={Modelling distinct failure mechanisms in composite materials by a combined phase field method},
  author={Zhang, Peng and Feng, Yaoqi and Bui, Tinh Quoc and Hu, Xiaofei and Yao, Weian},
  journal={Composite Structures},
  volume={232},
  pages={111551},
  year={2020},
  publisher={Elsevier}
}

@article{paggi2017revisiting,
  title={Revisiting the problem of a crack impinging on an interface: a modeling framework for the interaction between the phase field approach for brittle fracture and the interface cohesive zone model},
  author={Paggi, Marco and Reinoso, Jos{\'e}},
  journal={Computer Methods in Applied Mechanics and Engineering},
  volume={321},
  pages={145--172},
  year={2017},
  publisher={Elsevier}
}

@article{hansen2019phase,
  title={Phase-field modelling of interface failure in brittle materials},
  author={Hansen-D{\"o}rr, Arne Claus and de Borst, Ren{\'e} and Hennig, Paul and K{\"a}stner, Markus},
  journal={Computer Methods in Applied Mechanics and Engineering},
  volume={346},
  pages={25--42},
  year={2019},
  publisher={Elsevier}
}

@book{arnold1992ordinary,
  title={Ordinary differential equations},
  author={Arnold, Vladimir I},
  year={1992},
  publisher={Springer Science \& Business Media}
}

@article{francfort1998revisiting,
  title={Revisiting brittle fracture as an energy minimization problem},
  author={Francfort, Gilles A and Marigo, J-J},
  journal={Journal of the Mechanics and Physics of Solids},
  volume={46},
  number={8},
  pages={1319--1342},
  year={1998},
  publisher={Elsevier}
}

@article{karma2001phase,
  title={Phase-field model of mode III dynamic fracture},
  author={Karma, Alain and Kessler, David A and Levine, Herbert},
  journal={Physical Review Letters},
  volume={87},
  number={4},
  pages={045501},
  year={2001},
  publisher={APS}
}

@article{miehe2010thermodynamically,
  title={Thermodynamically consistent phase-field models of fracture: Variational principles and multi-field FE implementations},
  author={Miehe, Christian and Welschinger, Fabian and Hofacker, Martina},
  journal={International journal for numerical methods in engineering},
  volume={83},
  number={10},
  pages={1273--1311},
  year={2010},
  publisher={Wiley Online Library}
}

@article{bleyer2018phase,
  title={Phase-field modeling of anisotropic brittle fracture including several damage mechanisms},
  author={Bleyer, Jeremy and Alessi, Roberto},
  journal={Computer Methods in Applied Mechanics and Engineering},
  volume={336},
  pages={213--236},
  year={2018},
  publisher={Elsevier}
}

@article{zhang2019phase,
  title={Phase field modeling of fracture in fiber reinforced composite laminate},
  author={Zhang, Peng and Hu, Xiaofei and Bui, Tinh Quoc and Yao, Weian},
  journal={International Journal of Mechanical Sciences},
  volume={161},
  pages={105008},
  year={2019},
  publisher={Elsevier}
}

@article{yoshioka2021variational,
  title={Variational phase-field fracture modeling with interfaces},
  author={Yoshioka, Keita and Mollaali, Mostafa and Kolditz, Olaf},
  journal={Computer Methods in Applied Mechanics and Engineering},
  volume={384},
  pages={113951},
  year={2021},
  publisher={Elsevier}
}

@article{camacho1996computational,
  title={Computational modelling of impact damage in brittle materials},
  author={Camacho, Godofredo T and Ortiz, Michael},
  journal={International Journal of solids and structures},
  volume={33},
  number={20-22},
  pages={2899--2938},
  year={1996},
  publisher={Elsevier}
}

@article{bian2021novel,
  title={A novel phase-field based cohesive zone model for modeling interfacial failure in composites},
  author={Bian, Pei-Liang and Qing, Hai and Schmauder, Siegfried},
  journal={International Journal for Numerical Methods in Engineering},
  volume={122},
  number={23},
  pages={7054--7077},
  year={2021},
  publisher={Wiley Online Library}
}

@article{yue2025triple,
  title={A triple-damage model for fibrous composite material with intra-and inter-laminar decomposition, reduced-order-homogenization and phase field method},
  author={Yue, Jiajia and Yuan, Zifeng},
  journal={International Journal of Solids and Structures},
  pages={113490},
  year={2025},
  publisher={Elsevier}
}

@article{bian2024unified,
  title={A unified phase-field method-based framework for modeling quasi-brittle fracture in composites with interfacial debonding},
  author={Bian, Pei-Liang and Qing, Hai and Schmauder, Siegfried and Yu, Tiantang},
  journal={Composite Structures},
  volume={327},
  pages={117647},
  year={2024},
  publisher={Elsevier}
}

@article{vicentini2023phase,
  title={Phase-field modeling of brittle fracture in heterogeneous bars},
  author={Vicentini, Francesco and Carrara, Pietro and De Lorenzis, Laura},
  journal={European Journal of Mechanics-A/Solids},
  volume={97},
  pages={104826},
  year={2023},
  publisher={Elsevier}
}

@article{vicentini2025variational,
  title={Variational phase-field modeling of cohesive fracture with flexibly tunable strength surface},
  author={Vicentini, Francesco and Heinzmann, Jonas and Carrara, Pietro and De Lorenzis, Laura},
  journal={Journal of the Mechanics and Physics of Solids},
  pages={106424},
  year={2025},
  publisher={Elsevier}
}

@article{bourdin2025variational,
  title={A variational approach to fracture incorporating any convex strength criterion},
  author={Bourdin, Blaise and Marigo, Jean-Jacques and Maurini, Corrado and Zolesi, Camilla},
  journal={arXiv preprint arXiv:2506.22558},
  year={2025}
}

\end{document}